\documentclass[tradiabstract]{aa}
\usepackage{graphicx}
\usepackage{amsbsy}
\usepackage{amsmath}
\usepackage{natbib}
\usepackage{color}
\usepackage{txfonts}
\usepackage{url}
\usepackage{bm}

\DeclareMathSymbol{\varOmega}{\mathord}{letters}{"0A}
\DeclareMathSymbol{\varSigma}{\mathord}{letters}{"06}
\DeclareMathSymbol{\varPsi}{\mathord}{letters}{"09}
\DeclareMathSymbol{\varPhi}{\mathord}{letters}{"08}
\DeclareMathSymbol{\varGamma}{\mathord}{letters}{"00}

\begin{document}

\title{Anatomy of rocky planets formed by rapid pebble accretion \\ II.\
Differentiation by accretion energy and thermal blanketing}
\titlerunning{Anatomy of rocky planets formed by rapid pebble accretion II}

\author{Anders Johansen\inst{1,2}, Thomas Ronnet\inst{2}, Martin
Schiller\inst{1}, Zhengbin Deng\inst{1} \& Martin Bizzarro\inst{1}}
\authorrunning{Johansen et al.}

\institute{$^1$ Center for Star and Planet Formation, GLOBE Institute,
University of Copenhagen, \O ster Voldgade 5-7, 1350 Copenhagen, Denmark \\ $^2$
Lund Observatory, Department of Astronomy and Theoretical Physics, Lund
University, Box 43, 221 00 Lund, Sweden, \\e-mail:
\url{Anders.Johansen@sund.ku.dk}}

\date{}

\abstract{We explore the heating and differentiation of rocky planets that grow
by rapid pebble accretion. Our terrestrial planets grow outside of the ice line
and initially accrete 28\% water ice by mass. The accretion of water stops after
the protoplanet reaches a mass of $0.01\,M_{\rm E}$ where the gas envelope
becomes hot enough to sublimate the ice and transport the vapour back to the
protoplanetary disc by recycling flows. The energy released by the decay of
$^{26}$Al melts the accreted ice to form clay (phyllosilicates), oxidized iron
(FeO), and a water surface layer with ten times the mass of Earth's modern
oceans. The ocean--atmosphere system undergoes a run-away greenhouse effect
after the effective accretion temperature crosses a threshold of around 300 K.
The run-away greenhouse process vaporizes the water layer, thereby trapping the
accretion heat and heating the surface to more than 6,000 K. This causes the
upper part of the mantle to melt and form a global magma ocean.  Metal melt
separates from silicate melt and sediments towards the bottom of the magma
ocean; the gravitational energy released by the sedimentation leads to positive
feedback where the beginning differentiation of the planet causes the whole
mantle to melt and differentiate. All rocky planets thus naturally experience a
magma ocean stage. We demonstrate that Earth's small excess of $^{182}$W (the
decay product of $^{182}$Hf) relative to the chondrites is consistent with such
rapid core formation within 5 Myr followed by equilibration of the W reservoir
in Earth's mantle with $^{182}$W-poor material from the core of a planetary-mass
impactor, provided that the equilibration degree is at least 25\%-50\%,
depending on the initial Hf/W ratio. The planetary collision must have occurred
at least 35 Myr after the main accretion phase of the terrestrial planets.}


\maketitle

\section{Introduction}

Observations of young stars reveal that they are surrounded by very massive
protoplanetary discs of gas and dust, with a typical early-stage protoplanetary
disc containing several hundred Earth masses of solids in the form of
millimetre-sized pebbles \citep{Tychoniec+etal2018,Carrasco-Gonzalez+etal2019}.
These pebbles play an important role in the theory of planetesimal formation, as
the streaming instability \citep{Johansen+etal2007,BaiStone2010}, likely in
combination with weak pressure bumps that form in the turbulent gas
\citep{StollKley2016,MangerKlahr2018,Yang+etal2018,Schaefer+etal2020}, can
concentrate pebbles into dense filaments that undergo contraction and
gravitational collapse to form planetesimals with a characteristic size of
100 km \citep{Johansen+etal2015,Simon+etal2016,Nesvorny+etal2019}.

Pebbles can also drive planetary growth beyond the planetesimal stage. The
characteristic pebble size of 0.1--1 millimetre, inferred from observations of
protoplanetary discs \citep{Zhu+etal2019,Liu2019}, is in the right size
range for frictional energy dissipation to be significant during the scattering
by a growing protoplanet, which leads to capture and accretion of the pebble by
the protoplanet. This pebble accretion process has growth rates that are
significantly enhanced relative to the accretion rate of large planetesimals
\citep{JohansenLacerda2010,OrmelKlahr2010,LambrechtsJohansen2012}. High pebble
accretion rates are likely necessary to form planets akin to the gas giants and
ice giants in the Solar System where planetesimal accretion fails to form cores
within the lifetime of the protoplanetary disc
\citep{Bitsch+etal2015,JohansenBitsch2019,TanakaTsukamoto2019}.

The inner regions of the protoplanetary disc -- where the terrestrial planets
and super-Earths form -- are also traversed by the large population of drifting
pebbles. The traditional terrestrial planet formation theory was nevertheless
developed around a combination of planetesimal accretion and giant impacts
between the growing protoplanets \citep{KokuboIda1998} with water delivery by
rare impacts of icy planetesimals \citep{Raymond+etal2004}. The low value of the
excess of $^{182}$W relative to the non-radiogenic isotopes $^{183}$W and
$^{184}$W in Earth's mantle is interpreted as evidence that
Earth formed relatively slowly by accumulation of planetesimals and smaller
protoplanets. The isotope $^{182}$W is created by the decay of $^{182}$Hf with a
half life of approximately 8.9 Myr \citep{KleineWalker2017}. Since Hf is
lithophile and W is moderately siderophile, a low excess of $^{182}$W indicates
protracted core formation. However, it is not clear whether the Hf-W age of
Earth is a measure of the true accretion time or whether it is set mainly by a
later moon-forming giant impact.  Such a giant impact may have been the result
of an instability among a terrestrial planet system that originally contained
more planets \citep{Canup2012,QuarlesLissauer2015,Johansen+etal2021}. The main
core formation stage of Earth could therefore have lasted closer to 10 Myr,
followed by a moon-forming giant impact occurring more than 50 Myr later
\citep{Yin+etal2002,YuJacobsen2011}.

Rapid pebble accretion must consequently be explored as a serious candidate for
the formation mechanism of terrestrial planets, both around the Sun and around
other stars. \cite{Lambrechts+etal2019} demonstrated how the radial mass flux of
pebbles determines whether protoplanets in the inner regions of the
protoplanetary disc grow to Mars-sized protoplanets or continue to form
migrating chains of super-Earths. \cite{Johansen+etal2021} instead explored
intermediate pebble fluxes and found that the masses, orbits and volatile
budgets of the terrestrial planets in the Solar System (except Mercury) are
consistent with starting as protoplanets at the primordial ice line and
then growing and migrating to their current masses and orbits.

All four terrestrial planets in the Solar System are differentiated into a
metallic core dominated by iron and a silicate mantle where iron is only present
in the form of oxidized FeO \citep[see][for recent
reviews]{Carlson+etal2014,Tronnes+etal2019}. This widespread differentiation
implies that melting of the terrestrial planets was a natural consequence of
their accretion and early evolution -- and that all terrestrial planets go
through a magma ocean stage where the mantle was partially or wholly molten
\citep{Elkins-Tanton2012,SchaeferElkins-Tanton2018}. Impacts are traditionally
invoked to explain the formation of such magma oceans
\citep{MatsuiAbe1986a,MatsuiAbe1986b,RighterDrake1999,WadeWood2005,Rubie+etal2015a}.
The necessity for widespread magma oceans also implies that the primordial
atmospheres of the terrestrial planets were outgassed from the global magma
ocean as it recrystallized and expelled its incompatible volatiles. The
composition of the primordial atmospheres of terrestrial planets was thus
controlled by the overall budgets of the main volatiles H, C, and N combined
with the oxygen released by the dominant Fe-O-FeO reactions occurring in the
magma ocean \citep{SchaeferFegley2010,Ortenzi+etal2020}. The crystallization
partitioned a small fraction of the volatiles into solid crystals in the mantle
that can be compared to current constraints on carbon and hydrogen reservoirs in
the terrestrial mantle \citep{Elkins-Tanton2008}.

In Paper I of this series, we demonstrated how planetesimals forming outside
the ice line melt and desiccate through heat release by the decay of
$^{26}$Al. Energy from the decay of $^{26}$Al runs out after a few half-lives
($t_{1/2}=0.7\,{\rm Myr}$). We continue to explore here in Paper II the interior
evolution of planets that form by rapid pebble accretion. We add mass accretion
to the model and follow the heating and differentiation of the interior of the
growing planet, the outgassing of an early atmosphere that acts as thermal
blanketing, the emergence of a magma ocean and the differentiation of the planet
into a metal core and a silicate mantle. We show that rapid accretion by pebble
accretion combined with thermal blanketing from an outgassed atmosphere leads to
widespread melting of planets as they grow beyond approximately 1\% of an Earth
mass\footnote{We note that during revision of this paper a similar idea and
model were published by \cite{Olson+etal2022}.}. This way pebble accretion
naturally explains the emergence of the magma oceans that are needed to
understand differentiation and atmospheric outgassing.  We furthermore
demonstrate that the relatively small excess of $^{182}$W in Earth, when
compared to very early formed bodies such as Vesta, is consistent with a main
accretion phase lasting only 5 Myr followed by a moon-forming impact after
40--60 Myr that exchanged enough tungsten isotopes between the core of the
impactor and the mantle of Earth to lower the $^{182}$W excess to its observed
value.

The paper is organized as follows. In Section 2 we describe the modules of the
ADAP code related to partitioning of volatiles (H, C, O, and N) between the
core, mantle and atmosphere. In Section 3 we present the results of running the
ADAP code for the full accretion of a terrestrial planet. We focus particularly
on comparing the heating and differentiation of an Earth analogue and a Mars
analogue. In Section 4 we turn to calculating the evolution of the Hf and W
contents of the mantle and core of Earth before and after the moon-forming
impact. We show that the formation of Earth by rapid pebble accretion is
consistent with the low excess of $^{182}$W in the mantle of Earth for
significant impactor masses (higher than $0.2\,M_{\rm E}$) and high core-mantle
equilibration. We summarize our results in Section 5. Appendix A describes our
algorithm for calculating the temperature and pressure structure of the
outgassed atmosphere and the envelope of gas attracted from the protoplanetary
disc.

\section{ADAP code modules on volatile partitioning}

We describe here the aspects of the ADAP code related to distributing volatiles
between the core, mantle and atmosphere as well as the structure and cooling
rate of the outgassed atmosphere and accreted gas envelope. The interior heating
and composition evolution model of ADAP is described in Paper I. The atmospheric
composition is simplified to consist of the molecules H$_2$O, CO$_2$, H$_2$, CO,
O$_2$ and N$_2$. These molecules are assumed to be stored as H$_2$O, CO$_2$ and
N$_2$ within the mantle and core of the growing planet. The oxygen fugacity of
the magma ocean sets the speciation of H (H$_2$O or H$_2$) and C (CO$_2$ and CO)
as they outgas to the atmosphere. We include additionally the evaporation
under very hot conditions of non-volatile silicates from the magma ocean.
\begin{figure*}
  \begin{center}
    \includegraphics[width=0.75\linewidth]{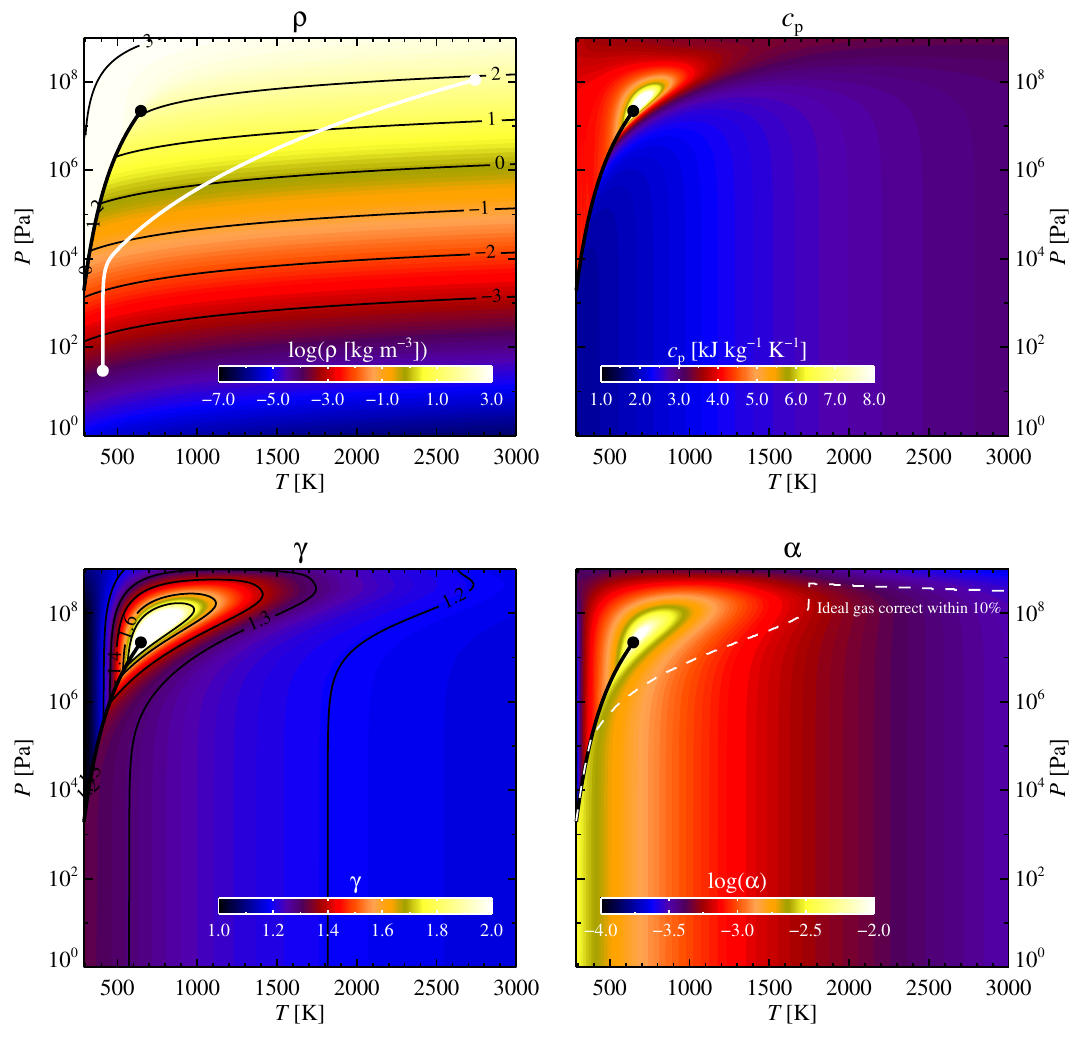}
  \end{center}
  \caption{Equation of state of water based on the IAPWS prescription
  (http://www.iapws.org/). The top left plot shows the density of the water as a
  function of temperature and pressure. The thick black line indicates the
  saturated vapour pressure and the black dot the critical point beyond which
  the distinction between liquid and vapour vanishes. The white line shows the
  structure of the atmosphere of our Earth analogue at the beginning of the
  differentiation phase ($t=3.5$ Myr). The top right plot shows the heat
  capacity at constant pressure and the bottom left the ratio of the specific
  heats, $\gamma=c_{\rm p}/c_{\rm v}$. Finally, the bottom right plot shows the
  thermal expansion coefficient $\alpha$, which is needed to calculate the
  adiabatic lapse rate of the planetary atmosphere. For an ideal gas law,
  $\alpha = (1/V) (\partial V / \partial T)_P = 1/T$. The dashed line indicates
  where the ideal gas law expression is correct within 10\% in the calculation
  of $\alpha$.}
  \label{f:eos_water}
\end{figure*}

The terrestrial planets grow massive enough inside the protoplanetary disc to
attract a significant hydrostatic envelope of hydrogen and helium gas
\citep{Stokl+etal2016,Johansen+etal2021}. This envelope adds additional thermal
blanketing to the surface and therefore must be included in the calculations.

\subsection{Atmosphere structure}

The surface pressure at the bottom of the atmosphere is calculated from
\begin{equation}
  P_{\rm sur} = \frac{g M_{\rm atm}}{4 \pi R^2} + P_{\rm env} \, .
  \label{eq:Psur}
\end{equation}
Here $g$ is the gravitational acceleration at the surface, $M_{\rm atm}$ is the
sum of the masses of the atmospheric components, $R$ is the planetary radius
(i.e.\ the distance from its centre to the solid or liquid surface, not
including the atmosphere) and $P_{\rm env}$ is the pressure at the bottom of
the hydrostatic gas envelope attracted from the protoplanetary disc. The
outgassed atmosphere transitions continuously to the attracted envelope in both
temperature and density at the pressure level $P=P_{\rm env}$. We assume that
the transition in mean molecular weight at the atmosphere-envelope transition
will create a narrow radiative zone at the transition, preventing mixing of
atmospheric constituents into the envelope. The complete structure of the
atmosphere-envelope system is thus given uniquely by the outwards-transported
luminosity, the mass of the atmosphere and the boundary conditions of the
envelope towards the protoplanetary disc. We assume further that the atmospheric
scale-height is so low that $g$ is constant and attains the value at the
planetary surface throughout the extent of the atmosphere $z$ (measured from the
surface).

The full scheme for finding the atmosphere-envelope structure and luminosity for
a given surface temperature is given in Appendix A. We use here an analytical
solution for the envelope structure based on the analysis presented in
\cite{PisoYoudin2014}. The outgassed atmosphere can not in general be assumed to
follow an ideal gas law, particularly when water vapour is under pressure and
temperature beyond the critical point. We therefore perform a full numerical
integration of the atmosphere from the surface to the atmosphere-envelope
transition.

The vertical temperature gradient of the atmosphere $\varGamma = -\partial
T/\partial z$ is constructed to follow either a dry adiabat,
\begin{equation}
  \varGamma = \frac{\alpha T g}{c_{\rm p}} \, ,
\end{equation}
or a more complex moist adiabat (including cloud particles) when the water
vapour in the atmosphere is saturated \citep{Leconte+etal2013}. Here $\alpha$ is
the thermal expansion coefficient, $T$ is the temperature and $c_{\rm p}$ is the
specific heat at constant pressure. Knowing the temperature lapse rate, the
pressure and density follow from hydrostatic equilibrium and an equation of
state. 

The equation of states for CO$_2$, H$_2$, CO, O$_2$ and N$_2$ are all assumed to
follow the ideal gas law; this yields a simple measure of the thermal expansion
coefficient as $\alpha = 1/T$. The specific heat capacities at constant pressure
of these gases are calculated from polynomial fits provided at the NIST
Chemistry WebBook (https://webbook.nist.gov/chemistry/). Since water undergoes
transitions both from vapour to liquid and from ideal gas to supercritical gas
phase, we adopted a parameterized gas law for water based on the IAPWS
prescription. We show our calculation of the equation of state of water in
Figure \ref{f:eos_water}. We create a look up table of the water density,
specific heats and thermal expansion coefficient at the set up of a simulation
and use this table for rapid, interpolated look-ups during runtime. Any
water amount in excess of the saturated vapour pressure at a given height over
the surface is converted to cloud particles with zero contribution to pressure,
but including the contribution of the liquid or frozen droplets to both density
and heat capacity.

\subsection{Luminosity carried through atmosphere and envelope}

We calculate the luminosity transported from the planetary surface through the
atmosphere and envelope as
\begin{equation}
  L_{\rm atm} = 4 \pi R^2 \sigma_{\rm SB} (T_{\rm sur}^4-T_{\rm atm}^4) \, ,
  \label{eq:Lpla}
\end{equation}
where $\sigma_{\rm SB}$ is the Stefan-Boltzmann constant, $T_{\rm sur}$ is the
temperature of the planetary surface, and $T_{\rm atm}$ is the temperature at
the bottom of the atmosphere. We iterate on this luminosity until the boundary
conditions of the protoplanetary disc at the Hill sphere are fulfilled (see
Appendix A). This approach will nevertheless lead to very high surface
temperatures (up to 10,000 K or more). This is not realistic, since the critical
point of SiO is approximately 6,600 K \citep{XiaoStixrude2018}, above which the
magma ocean has no distinct liquid and vapour phases. Pebbles would thus not
reach the surface and the accretion energy would be released where the pebbles
are dissolved higher up the atmosphere; this will prevent burial of the
accretion heat and thus reduce the temperature at the surface. We therefore
distribute part of the pebble accretion heat at the bottom of the envelope,
rather than in the surface layer, when the temperature of the lower atmosphere
is above a threshold value of $T_{\rm atm,max}=3{,}000\,{\rm K}$. We do not
release any silicate vapour in the hydrostatic envelope above the atmosphere
\citep{Brouwers+etal2018}; this assumption is valid for the relatively modest
temperatures reached in the base of the envelope (see Figure
\ref{f:atmosphere_structure_Earth}).

The fraction of accretion energy dissipated in the bottom of the envelope is set
to
\begin{equation}
  f_{\rm acc,env} = \tanh [ (T/T_{\rm atm,max})^{10} ] \, .
\end{equation}
This ensures a transition width of a few hundred K from 100\% energy release in
the surface layer ($f_{\rm acc,env}=0$) to 100\% energy release at the bottom of
the envelope ($f_{\rm acc,env}=1$). The heating of the surface is thus reduced
to $L_{\rm sur} = (1-f_{\rm acc,env}) L_{\rm acc}$ and the bottom of the
envelope carries the total luminosity
\begin{equation}
  L_{\rm env} = L_{\rm atm} + f_{\rm acc,env} L_{\rm acc} \, .
\end{equation}

After the dissipation of the protoplanetary disc, the external boundary
condition no matter bears physical meaning. Instead, the luminosity known from
equation (\ref{eq:Lpla}) will then define the temperature $T_{2/3}$ at optical
depth $\tau=2/3$ through
\begin{equation}
  L = 4 \pi R_{\rm pho}^2 \sigma_{\rm SB} T_{2/3}^4  \, ,
\end{equation}
with $R_{\rm pho}$ denoting the photosphere radius. In Appendix A we describe
how this is calculated from the knowledge of the envelope opacity and the
luminosity.

\begin{figure*}
  \begin{center}
    \includegraphics[width=0.8\linewidth]{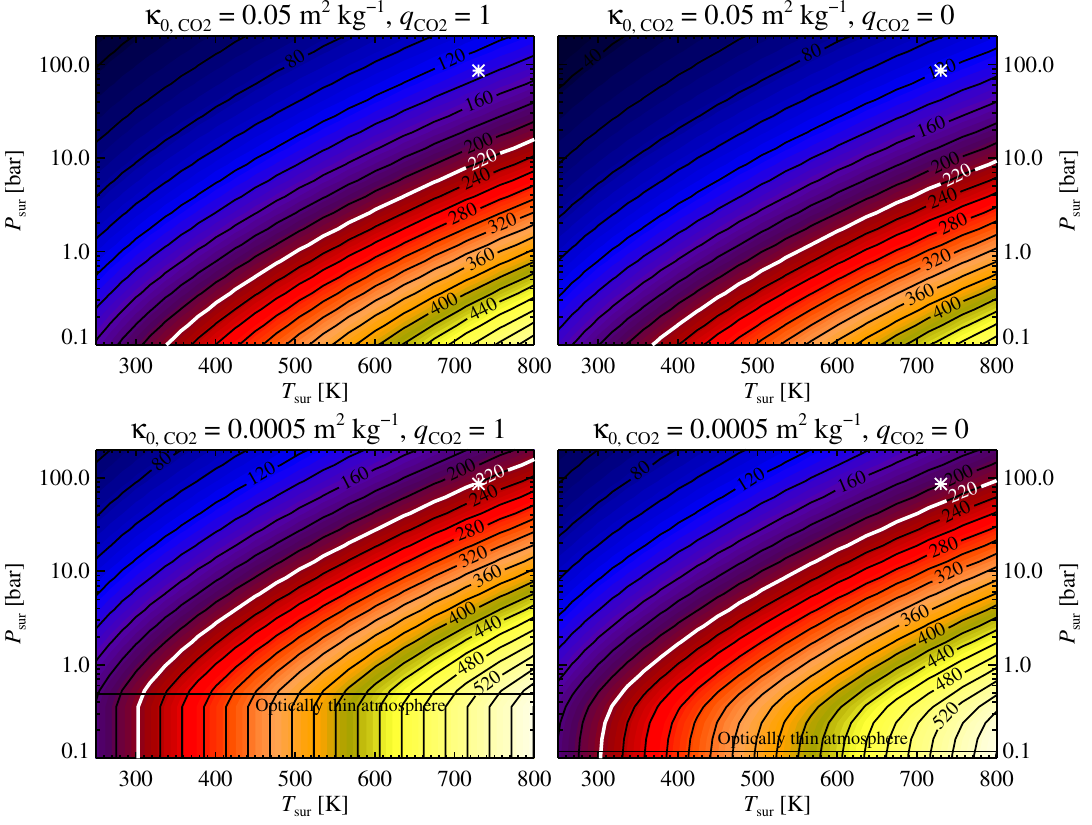}
  \end{center}
  \caption{Photospheric temperature of a Venus analogue (in terms of
  mass and radius), for a range of surface temperatures ($x$-axis) and surface
  pressure ($y$-axis). The atmospheric composition is assumed to be 100\%
  CO$_2$.  The top plots utilize a high CO$_2$ opacity, while the bottom plots
  are calculated using our nominal value that is 100 times lower \citep[see
  discussion in][]{Elkins-Tanton2008}. The left plots use the nominal linear
  dependence of the opacity on the pressure, while the right plots show the
  difference when using a pressure-independent opacity instead. Our atmosphere
  model becomes invalid below the horizontal black line, which indicates the
  transition to an optically thin atmosphere. The effective temperature of Venus
  (220 K, when taking into accounts its relatively high current opacity) is
  indicated with a thick line and the temperature and surface CO$_2$ pressure
  with an asterisk. We choose therefore the low opacity value and linear
  pressure dependence to best reproduce the atmosphere of Venus in our
  calculations.}
  \label{f:luminosity}
\end{figure*}

In the absence of a gas envelope, for example after its removal by XUV
radiation, the photospheric temperature level $T_{2/3}$ is reached in the
outgassed atmosphere. In that case we integrate
\begin{equation}
  {\rm d}\tau = \kappa(P) \rho {\rm d}z
\end{equation}
from $\tau=0$ to $\tau=2/3$. We use an opacity law $\kappa=\kappa_0
(P/P_0)^q$ that is proportional to the local pressure $P$ \citep[$q=1$,
see][]{Badescu2010}, with $P_0=1.013 \times 10^5$ Pa and $\kappa_0 = 1.0\,{\rm
m^2\,kg^{-1}}$ for water \citep{MatsuiAbe1986a,Nakajima+etal1992,KollCronin2019}
and $\kappa_0 = 0.0005\,{\rm m^2\,kg^{-1}}$ for CO$_2$. The other
atmospheric components are assumed to carry insignificant opacities.
\cite{Elkins-Tanton2008} discusses whether a low value (our nominal choice) or a
high value for the opacity of CO$_2$ is most realistic. We demonstrate in Figure
\ref{f:luminosity} the photospheric temperature of a Venus analogue planet with
a pure CO$_2$ atmosphere and a range of surface pressures and surface
temperatures.  We also vary the opacity law from linear with $P$ (the nominal
choice) to constant. We find the best value for the surface temperature of Venus
to be represented by $\kappa_0 = 0.0005\,{\rm m^2\,kg^{-1}}$ for CO$_2$ with a
linear opacity law. This choice of a low CO$_2$ opacity agrees with the
CO$_2$ atmosphere models of \cite{Wordsworth2015} and \cite{Salvador+etal2017}.

\begin{figure}
  \begin{center}
    \includegraphics[width=0.8\linewidth]{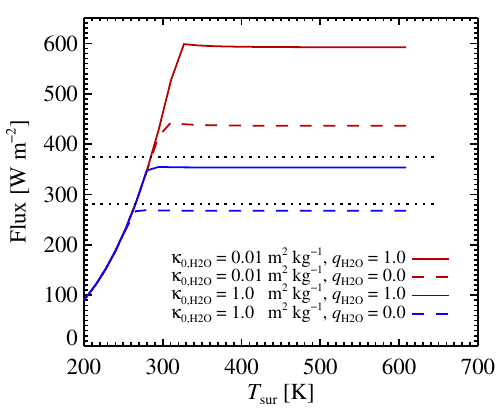}
  \end{center}
  \caption{The thermal flux escaping the top of a pure water atmosphere for
  a range of surface temperatures. The run-away greenhouse effect manifests
  itself as a plateau in the outgoing flux: there is no stable thermal
  equilibrium when the external or internal heating is higher than this critical
  flux $F_{\rm crit}$.  We mark the \cite{Goldblatt+etal2013} result of a full
  frequency-dependent radiative transfer 1-D model ($F_{\rm crit}=282\,{\rm
  W\,m^{-2}}$) and the \cite{Leconte+etal2013} result of a 3-D climate model
  ($F_{\rm crit}=375\,{\rm W\,m^{-2}}$) with dotted lines. The grey opacity
  parametrization of \cite{MatsuiAbe1986a}, with $\kappa_0 = 0.01\,{\rm
  m^2\,kg^{-1}}$ and linear pressure dependence (full red line), results in a
  too high critical flux value. We find a better match when adopting a two
  orders of magnitude higher opacity, $\kappa_0 = 1.0\,{\rm m^2\,kg^{-1}}$.}
  \label{f:luminosity_water}
\end{figure}

The water opacity influences the transition from stable atmosphere to
run-away greenhouse effect. \cite{Ingersoll1969} demonstrated that a pure water
atmosphere in equilibrium with a surface ocean will obtain a maximum cooling
rate at an effective temperature of around 300 K, beyond which any increase in
the surface temperature will not lead to additional cooling \citep[see
also][]{Hamano+etal2013,Kopparapu+etal2013}. We test the run-away greenhouse
effect for an Earth analogue in Figure \ref{f:luminosity_water}. Using the
opacity model of \cite{MatsuiAbe1986a}, with $\kappa_0 = 0.01\,{\rm
m^2\,kg^{-1}}$ at one atmosphere pressure and linear pressure dependence, the
transition to run-away greenhouse effect happens at a surface temperature of
around 320 K. \cite{Goldblatt+etal2013} demonstrated using frequency-dependent
radiative transfer models that the transition happens at an outgoing flux of 282
${\rm W}\,{\rm m}^{-2}$. On the other hand, \cite{Leconte+etal2013} found a
limiting flux of 375 ${\rm W}\,{\rm m}^{-2}$.  We match this range of values
best with a much higher water opacity level with $\kappa_0 = 1.0\,{\rm
m^2\,kg^{-1}}$ and therefore adopt this (with linear pressure dependence) as our
nominal value. We experiment in Figure \ref{f:core_fraction_t} with lower
values of the water opacity to assess how this affects the differentiation of
the planet.

\subsection{Hydrogen opacity}

We assume that the H$_2$ molecule carries zero opacity in our calculations.  At
high densities of CO$_2$ and H$_2$, collision-induced absorption can raise the
opacity of the H$_2$ molecule to significant values \citep{Karman+etal2019}.  We
therefore performed a numerical experiment with an H$_2$ opacity equal to the
(high) H$_2$O opacity. We found the H$_2$ opacity to have very little effect on
the mass onset for differentiation (see effect of water opacity in Figure
\ref{f:core_fraction_t}). This is likely due both to the high water opacity,
which makes other opacity sources insignificant, and to the fact that the flux
threshold for the run-away greenhouse effect depends only weakly on the opacity
(see Figure \ref{f:luminosity_water}).

\subsection{Atmospheric escae}

We describe our energy-limited approach to remove the outgassed atmosphere and
the envelope by XUV irradiation in Paper III.

\subsection{Water evaporation}

We outgas a water vapour atmosphere from surficial and underground water
(evaporation) or ice (sublimation) by calculating the saturated vapour pressure
of water, $P_{\rm sat}$, at the equilibrium temperature $T_{\rm equ}$ using the
IAPWS equation of state. In equilibrium, the water mixing ratio $X_{\rm H_2O}$
at the surface must fulfil
\begin{equation}
  P_{\rm sat} = X_{\rm H_2O} P_{\rm sur} \, .
\end{equation}
Using equation (\ref{eq:Psur}) and an expression for the mixing ratio, the right
hand side can further be expanded as
\begin{equation}
  P_{\rm sat} = \frac{M_{\rm sat}/\mu_{\rm H_2O}}{\sum_i M_i / \mu_i} \left(
  \frac{g \sum_i M_i}{4 \pi R^2} + P_{\rm env} \right) \, .
  \label{eq:Psat2}
\end{equation}
Here $M_i$ is the mass of the $i$th atmospheric component, with $M_i = M_{\rm
sat}$ for water. We solve Equation (\ref{eq:Psat2}) for the saturated water mass
$M_{\rm sat}$ by transforming the equation to a second order polynomial.

When the surface layer is ice or water, then we evaporate or condense $f=0.1$
times the difference between the current atmospheric water mass and the
saturated water mass. However, we limit the outgassed mass to be $f$ times the
total water mass remnant in the planet. This ensures that the atmosphere and the
surface reach equilibrium within a few ten time-steps. We have compared a
simulation with $f=0.001$ to the nominal simulation with $f=0.1$ and found
nearly identical results; hence the vapour equilibrium is obtained on a short
enough time-scale that disequilibrium effects do not affect the results.

We check also whether the local pressure is higher than the saturated vapour
pressure below the surface. Deep boiling of water occurs whenever
$P(r)<P_{\rm sat}(r)$. This is particularly true when the water temperature is
above the critical temperature in which the liquid phase ceases to exist and the
saturated vapour pressure is formally infinite. We assume the deep boiling leads
to vapour bubble formation and outgassing at the local surface. In that case, we
require that the partial pressure of water in the atmosphere to equal the deep
boiling saturated pressure. This typically leads to outgassing of the entire
water contents of the protoplanet.

\subsection{Evaporation of silicates}

At surface temperatures above approximately 2,000 K, the silicate magma
ocean equilibrates with a significant vapour pressure in the atmosphere. The
vapour pressure over the magma ocean consists of a mixture of molecules such as
Mg, SiO, O and O$_2$ \citep{SchaeferFegley2004}. We follow the approach of
\cite{MisenerSchlichting2022} and define a temperature-dependent effective
silicate vapour pressure over the magma ocean, with a single mean molecular
weight. We choose SiO as the representative molecule of the silicate vapour; we
refer to \cite{MisenerSchlichting2022} for a discussion of using a slightly
lighter value to reflect the mixture of species in the vapour. We set the
critical temperature to 6,600 K \citep{XiaoStixrude2018}, above which we freeze
the saturation pressure at the critical pressure level of $P_{\rm crit}=1.9
\times 10^4\,{\rm bar}$. We use an ideal gas equation of state for the SiO in
the atmosphere; this is a valid approximation because any SiO pressure in excess
of the critical pressure is not allowed to leave the magma ocean. We include
formation of SiO clouds in cooler layers above the surface, using a scheme
similar to water clouds. We ignore any latent heat release in SiO cloud
formation.

As discussed above, we limit the surface temperature to approximately 3,000 K to
reflect pebble destruction in the atmosphere. We motivate this choice here,
because at such temperatures silicates will dominate the composition of the
atmosphere. Silicate clouds will thus form high up in the atmosphere and we
assume that pebbles colliding with these clouds will be destroyed and hence
release their accretion energy at the base of the envelope rather than in the
magma ocean.

\subsection{Carbon and nitrogen release}

We treat carbon and nitrogen as trace species that do not contribute to the bulk
planetary mass (although both species add mass and pressure to the atmosphere
and carbon carries also atmospheric opacity). We set the carbon contents of the
accreted material to 2,000 ppm up to a maximum carbon amount of $4 \times
10^{-4}$ $M_{\rm E}$. This follows the model of volatile accretion by pebble
snow from \cite{Johansen+etal2021} where the carbon in organics is sublimated
between 325 K and 425 K and the remaining carbon in graphite burns at a
temperature of 1,100 K \citep{GailTrieloff2017} and subsequently diffuses
back to the protoplanetary disc within ultravolatile molecules such as CH$_4$
and CO. We assume that the carbon forms carbonates in the mantle after the
melting of the ice and that these carbonates decompose at $T_{\rm dec}=1000$ K,
releasing carbon into the mantle as CO$_2$.  For the nitrogen, we assume that it
resides mainly in the organics and is released, together with the carbon,
successively between 325 K and 425 K \citep{Nakano+etal2003,GailTrieloff2017}.
We release the nitrogen into the mantle at $T_{\rm dec}=1000$ K, similarly to
the carbon, and limit the total nitrogen amount to the current atmosphere of
Earth, $M_{\rm N_2}=3.87 \times 10^{18}$ kg. Since atmospheric nitrogen does not
contribute to the heating of the surface, the total amount of nitrogen in the
model is arbitrary and we are mainly interested in how the nitrogen partitions
between core, mantle and atmosphere -- this partitioning is discussed further in
Paper III.

\subsection{Water dissolution in magma}
\label{s:dissolution}

For the dissolution of water in magma we calculate first the total mass of
molten silicates $M_{\rm mag}$. We then calculate the equilibrium mass fraction
of water in the melt from \cite{MatsuiAbe1986a,MatsuiAbe1986b},
\begin{equation}
  X_{\rm wat} = 2.08 \times 10^{-6} (P_{\rm H_2O}/{\rm Pa})^{0.54} \, .
  \label{eq:Xwat}
\end{equation}
This equilibrium describes the reaction between O$^{2-}$ ions in the magma melt
and gaseous H$_2$O to form dissolved OH$^{-}$ radicals
\citep{Iacono-Marziano+etal2012,Ortenzi+etal2020}. The reaction of two OH$^{-}$
radicals to reform H$_2$O explains the square-root like pressure-dependence in
equation (\ref{eq:Xwat}).  We calculate $X_{\rm wat}$ from the water surface
pressure $P_{\rm sur,H_2O}$ and then we transfer 0.1 times the difference
between the equilibrium value of the water mass in the silicates and the actual
value between the atmosphere and the silicates. The magma ocean is assumed
to equilibrate with the atmosphere when the magma is present at higher levels
than 80\% of the current protoplanet radius. We define here magma as those
spherical shells that have a melt fraction higher than $\varPhi_{\rm
melt}=0.5$.

\subsection{CO$_2$ and N$_2$ dissolution in magma}

For partitioning of CO$_2$ between magma and atmosphere, we use the equilibrium
atmospheric partial pressure from \cite{Papale1997} and \cite{Hirschmann2012},
\begin{equation}
  X_{{\rm CO}_2} = 4.4 \times 10^{-12} (P_{\rm CO_2}/{\rm Pa}) \, .
\end{equation}
This equilibrium describes the reaction between O$^{2-}$ ions in the magma
melt and gaseous CO$_2$ to form dissolved carbonate CO$_3^{2-}$
\citep{Iacono-Marziano+etal2012,Ortenzi+etal2020}. The CO$_2$ solubility
changes with pressure and temperature at depth of the magma ocean, first
increasing for the terrestrial magma ocean down to a few hundred kilometers and
then falling until diamond condenses out at depths below 500 km
\citep{Hirschmann2012,Armstrong+etal2019}, corresponding to a pressure between
15 and 20 GPa. The concentration of carbon in the magma ocean is nevertheless
set by the equilibrium concentration of CO$_2$ at the surface interface, as this
value is maintained throughout the magma ocean by convection. Lower solubility
at depth implies that the molecular storage of carbon changes but the
concentration does not change as long as the diamond crystals do not grow to
large enough sizes to sediment. We note that \cite{Elkins-Tanton2008} seem to
use a much higher value for the solubility of carbon in magma that is close to
the solubility value for water (their equations 3 and 4).

For N$_2$, we follow \cite{Sossi+etal2020} and set the equilibrium concentration
in the magma ocean according to
\begin{equation}
  X_{{\rm N}_2} = 6.11 \times 10^{-13} (P_{\rm N_2}/{\rm Pa}) \, .
\end{equation}
This expression is valid for relatively oxidized conditions, while extremely
reduced conditions not encountered during the differentiation of Earth and Mars
(but maybe relevant for Mercury) are characterized by very high chemical
dissolution values \citep{Libourel+etal2003}. We refer to
\cite{Lichtenberg+etal2021} for a recent overview of the different dissolution
laws and their applicability ranges\footnote{The probed pressure range of
the dissolution laws extends to values way above the maximum pressures obtained
in our models, see Figure \ref{f:Psur_t}.}.

\subsection{Hydrogen, carbon and nitrogen in the core}

We consider the core and the mantle to be continuously connected through a
partition coefficient $D = C_{\rm cor}/C_{\rm man}$ where $C_{\rm cor}$ is the
concentration of the species in the core and $C_{\rm man}$ is the concentration
of the species in the magma mantle.  This yields an equilibrium mantle
concentration of
\begin{equation}
  C_{\rm man} = \frac{M_{\rm vol}}{D M_{\rm cor} + M_{\rm man}} \, ,
  \label{eq:Cman}
\end{equation}
where $M_{\rm vol}$, $M_{\rm man}$ and $M_{\rm cor}$ are the instantaneous
masses of the volatile species under consideration, the mantle and the core,
respectively. Volatiles may in reality not be able to partition continuously 
between metal and silicates subsequent to their initial transport to the core.
Assuming therefore that volatiles partition only between metal and silicates
during their descent through the mantle yields the mantle equilibrium
concentration
\begin{eqnarray}
  \dot{C}_{\rm man} &=& \frac{{\rm d}}{{\rm d}t} \left( \frac{M_{\rm
  vol,man}}{M_{\rm man}} \right) \nonumber \\ &=& \frac{\dot{M}_{\rm
  vol,man}}{M_{\rm man}} - \frac{M_{\rm vol,man}}{M_{\rm man}^2} \dot{M}_{\rm
  man} \nonumber \\ &=& \frac{\dot{M}_{\rm vol}}{M_{\rm man}} - D \frac{M_{\rm
  vol,man}}{M_{\rm man}} \frac{\dot{M}_{\rm cor}}{M_{\rm man}} - \frac{M_{\rm
  vol,man}}{M_{\rm man}^2} \dot{M}_{\rm man} \nonumber \\ &=& \frac{\dot{M}_{\rm
  vol}}{M_{\rm man}} - D C_{\rm man} \frac{\dot{M}_{\rm cor}}{M_{\rm man}} -
  C_{\rm man} \frac{\dot{M}_{\rm man}}{M_{\rm man}} \, .
\end{eqnarray}
This differential equation has an equilibrium solution, with $\dot{C}_{\rm
man}=0$, given by the mantle concentration
\begin{equation}
  C_{\rm man} = \frac{\dot{M}_{\rm vol}}{D \dot{M}_{\rm cor}+\dot{M}_{\rm man}}
  \, .
  \label{eq:Cman2}
\end{equation}
When $D$ is assumed to be a constant that does not depend on temperature and
pressure and all mass accretion rates are proportional to the instantaneous mass
(so that $\dot{M}_i = (M_i/M) \dot{M}$ for all components), then equation
(\ref{eq:Cman2}) transforms to the same as expression as when assuming
continuous partitioning between core and mantle (equation \ref{eq:Cman}). Since
we consider constant $D$ in the simulations in order to probe the range of
possible values for the partition coefficients, we apply equation \ref{eq:Cman}
in each time-step to determine the equilibrium concentrations of volatiles in
the mantle and the core.

Hydrogen, carbon and nitrogen are all siderophile elements with an affinity to
enter metallic melt over silicate melt. For hydrogen, we base our partition
coefficients between metal and silicates on \cite{KuramotoMatsui1996} and
\cite{Li+etal2020} and take a base value of $D_{\rm H}=5$. For carbon,
\cite{Fischer+etal2020} provided fits for $D_{\rm C}$ over a range of pressure
and temperatures. Since the partition coefficient of carbon depends so strongly
on the planetary mass, we experiment with a range of values of $D_{\rm C}$
around a base value of 300. Nitrogen does not affect the evolution of our
planets and is used mainly as a tracer for comparing observed atmospheres with
our model atmospheres.  Nitrogen partition coefficients additionally show strong
dependencies on the oxygen fugacity of the mantle \citep{Grewal+etal2019}. We
therefore experiment with partition coefficients for N between $D_{\rm N}=1$ and
$D_{\rm N}=100$ to bracket the range of realistic values, with a base value of
$D_{\rm N}=10$.

The core is assumed to equilibrate with the magma ocean when the lowest molten
molten mantle is less than 10\% of the protoplanet radius away from the
core-mantle boundary.

\subsection{Silicon and oxygen in the core}

We do not partition the elements Si and O to the metal, but we note that
\cite{Badro+etal2015} found that early oxidizing conditions following by an
increasing reduction of the accreted material, similar to what we propose for
the pebble accretion model in Paper I, is needed to reconcile the core density
with the inclusion of Si and O.

\subsection{Outgassing of H and C from the magma ocean}

We base the model for speciation of H and C from the magma ocean to the
atmosphere on the approach of \cite{Ortenzi+etal2020}. We assume that the magma
ocean operates as an oxygen buffer that effectively absorbs or emits oxygen to
the atmosphere at the level for balance between Fe, O$_2$ and FeO in the magma.
We calculate the oxygen fugacity $f$O$_2$ (effectively the partial pressure of
O$_2$ in the atmosphere measured in bars) relative to the iron-w\"ustite buffer
(IW). The value of fO2 at the IW buffer is taken from \cite{Pack+etal2011}, we
refer the reader also to \cite{Hirschmann2021}. We use IW+2 for the outgassing
on Earth and Venus, since these planets achieved relatively oxidized upper
mantles by the reaction FeO + (1/4) O$_2$ $\rightleftharpoons$ FeO$_{1.5}$ in
the lower mantle \citep[converting Fe$^{2+}$ to Fe$^{3+}$,
see][]{Armstrong+etal2019}, while for Mars we use the nominal IW-2 for magma
ocean that is poor in Fe$^{3+}$ \citep{WadeWood2005}. This leads to outgassing
of mainly CO$_2$ and H$_2$O for Earth and Venus, while Mars outgasses a reduced
atmosphere dominated by CO and H$_2$. A more realistic model would consider
the evolution of the oxygen fugacity with increasing depth of the magma ocean
\citep{Hirschmann2022} as well as oxidation due to hydrogen escape from the
atmosphere \citep{Wordsworth+etal2021}, but we opt for simplicity here to
maintain a constant oxygen fugacity relative to the IW buffer throughout
planetary accretion, early atmospheric evolution and mass loss phases.

\subsection{Ingassing of ultra-volatiles from the protoplanetary disc}

It has been proposed that part of the terrestrial reservoirs of water and noble
gases was sourced directly from the protoplanetary gas disc by ingassing
\citep{Wu+etal2018,WilliamsMukhopadhyay2019}. However, in our model the
outgassed atmosphere of high mean molecular weight separates the solar
protoplanetary disc from the magma ocean. The degree to which hydrogen and noble
gases can penetrate the mean molecular weight barrier between the outgassed
atmosphere and the gas attracted from the protoplanetary disc is unknown and we
therefore do not include any ingassing of ultra-volatiles from the
protoplanetary disc in our model. Efficient mass exchange between envelope
and atmosphere would likely drain the atmosphere of any free oxygen as O$_2$
readily combines with H$_2$ to form H$_2$O, thereby robbing FeO in the magma
ocean of its oxygen and adding metallic Fe to the core. We consider this an
unrealistic scenario given the 8\% FeO content of the terrestrial mantle
\citep[Paper I, see also][]{RighterO'Brien2011}. In order to understand the
effect of the hydrogen-helium envelope on the differentiation of the planet in
isolation, we nevertheless perform an additional numerical experiment where we
suppress the outgassed atmosphere and consider the hydrogen-helium envelope to
touch the magma ocean directly (see Figure \ref{f:core_fraction_t}).
\begin{figure*}
  \begin{center}
    \includegraphics[width=0.9\linewidth]{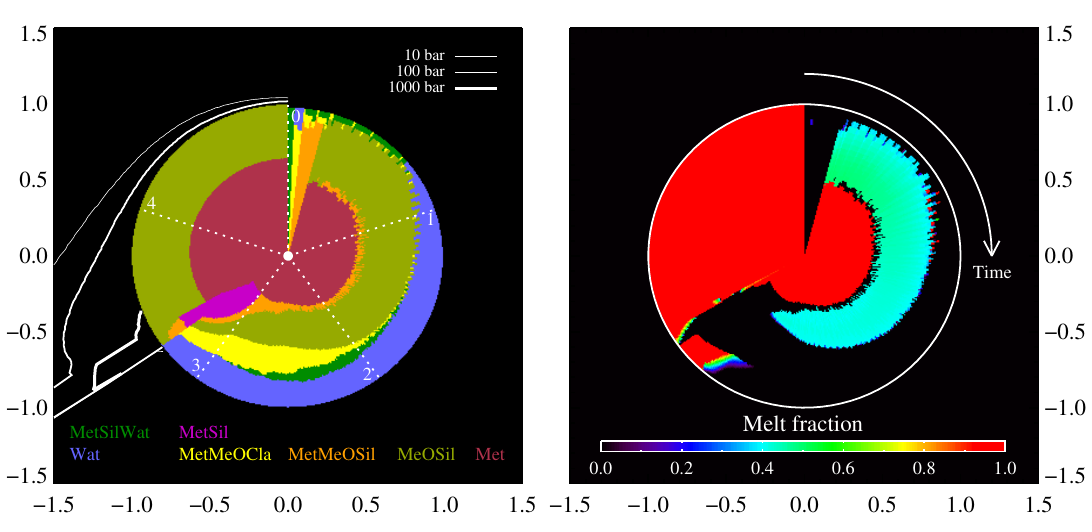}
  \end{center}
  \caption{Interior structure (left) and interior melting degree (right) of our
  Earth analogue shown with time from 0 to 5 Myr represented by the angle.
  The distance from the centre is normalized by the instantaneous planetary
  radius. Heating due to decay of $^{26}$Al initially leads to melting of the
  primitive MetSilWat material to form clay beneath a layer of ice. Further
  heating leads to full interior differentiation within 0.3 Myr. As the
  radiogenic energy runs out, the continued accretion leads to the emergence of
  a thick clay mantle overlaid by a massive surface ocean after 3 Myr. The
  accretion energy finally becomes high enough, after approximately 3.4 Myr, for
  a run-away greenhouse effect in the ocean-atmosphere system to heat the
  surface to above the melting temperature of silicates; this leads to a
  wholesale differentiation of the planet into a silicate mantle and a metal
  core. The contour lines show the 10, 100 and 1,000 bar pressure levels in
  the outgassed atmosphere.}
  \label{f:planet_structure_Earth_circle}
\end{figure*}
\begin{figure}
  \begin{center}
    \includegraphics[width=0.9\linewidth]{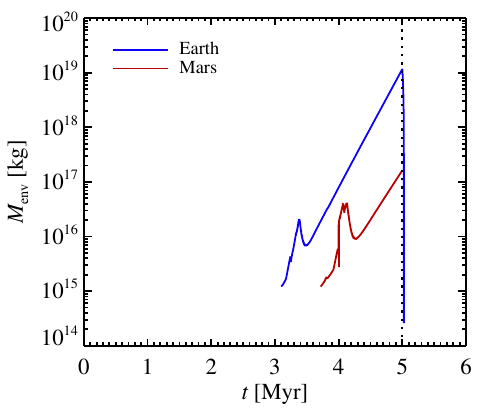}
  \end{center}
  \caption{The mass of the envelope as a function of time for Earth and Mars.
  Mars attracts far less gas from the protoplanetary disc due to its lower mass.
  The jump in envelope mass at 3.4 Myr for Earth and at 4.0 Myr for Mars is
  associated with the differentiation of the planets and the outgassing of the
  massive H$_2$O and CO$_2$ atmosphere that is efficient at trapping the
  accretion heat. The escape of the envelope by XUV irradiation from the star
  after the end of accretion at 5 Myr is extremely rapid due to the low envelope
  masses.}
  \label{f:Menv_t}
\end{figure}

\subsection{Partitioning of H$_2$O and CO$_2$ between magma and solids}

As the magma ocean crystallizes, its volatile contents are distributed between
the remaining magma and the newly formed solid crystals. Volatiles are typically
incompatible, meaning that they partition strongly towards the liquid magma. An
equilibrium is obtained when $C_{\rm sol} = k C_{\rm mag}$, with the equilibrium
concentration of volatiles in the solids $C_{\rm sol}$ a factor $k$ (generally
$\ll 1$) times the concentration of volatiles in the magma $C_{\rm mag}$. We use
pressure-dependent values of $k$ for hydroxyl (OH$^-$) and carbon from
\cite{Elkins-Tanton2008}. The carbon partitioning is valid for relatively
oxidized mantles; we refer to \cite{Ortenzi+etal2020} for an
oxygen-fugacity-dependent expression. However, for simplicity we choose to use
here the expressions from \cite{Elkins-Tanton2008}. For nitrogen, we use the
same solid-melt partition coefficient as for carbon. As the total magma mass
decreases, the concentration of volatiles in the remaining magma increases and
this leads to extensive transfer of volatiles from the mantle to the atmosphere
and the core.
\begin{figure*}
  \begin{center}
    \includegraphics[width=0.45\linewidth]{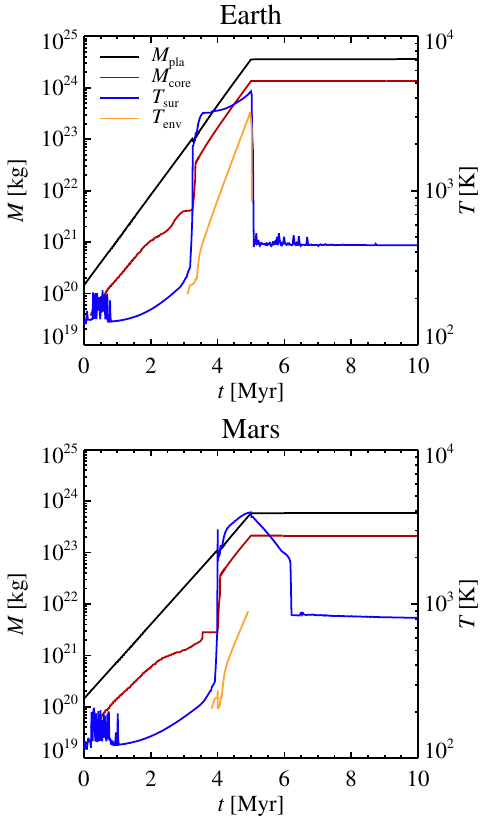}
    \includegraphics[width=0.45\linewidth]{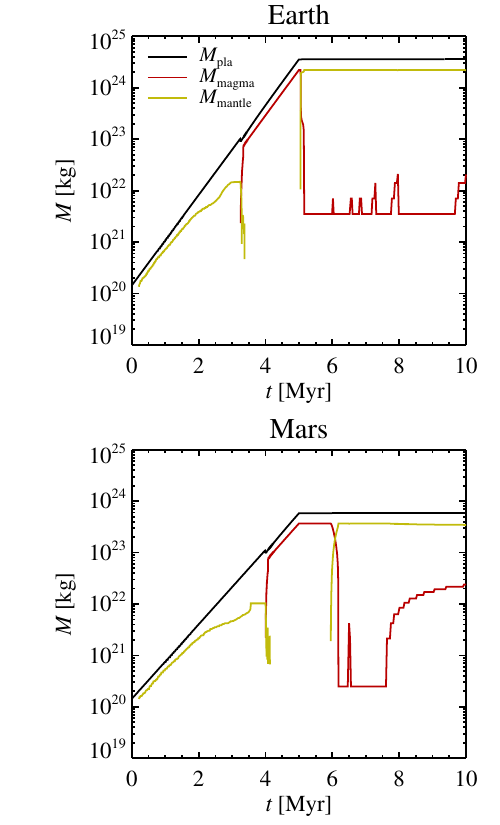} \\
  \end{center}
  \caption{Temporal evolution of mass reservoirs and characteristic
  temperatures. Left panels: Total mass and core mass (left axis) and surface
  temperature and envelope temperature (right axis) as a function of time for
  our Earth and Mars analogues. Right panels: Mass of the planet, the magma
  ocean and the solid mantle as a function of time. Within the first 3 Myr, the
  decay of $^{26}$Al provides enough energy to drive differentiation into an
  early core and mantle. The mantle temperature is kept below the liquidus
  temperature by efficient convective heat transport. A global magma ocean forms
  when the accretion energy blanketed by the outgassed atmosphere triggers a
  run-away greenhouse effect.  The surface temperature then rises to 4,500 K.
  The global magma ocean undergoes separation of metal melt from silicate melt,
  leading to the differentiation into a core and a mantle. The surface
  temperature falls steeply after the protoplanetary disc dissipates, as the
  accretion energy vanishes and the hydrostatic gas envelope escapes by stellar
  XUV irradiation. The surface temperature is nevertheless much higher than the
  effective temperature because of the greenhouse effect in the massive
  outgassed atmosphere. The magma ocean crystallizes rapidly after the end of
  the accretion phase, as its thermal energy is carried upwards by convection
  and transported through the atmosphere. The base of the magma ocean
  nevertheless experiences some remelting due to energy transfer from the still
  molten core (this is visible as spikes in the magma contents after 5 Myr).}
  \label{f:mass_t}
\end{figure*}

\section{Heating and differentiation of growing planets}

We run simulations for Earth and Mars analogues using for simplicity exponential
growth curves. We grow Mars to its modern mass, while Earth is grown only to
$0.6\,M_{\rm E}$ because of the later impact with Theia. Theia is constructed to
grow to $0.4\,M_{\rm E}$ in the pebble accretion model of
\cite{Johansen+etal2021}. We present additionally results for Venus in Paper
III, focusing mainly on comparing the atmospheric outgassing and escape between
Venus, Earth and Mars.

\subsection{Temporal evolution of composition and melting degree}

We show the evolution of our Earth analogue composition and melting degree in
Figure \ref{f:planet_structure_Earth_circle}. The activity of $^{26}$Al is high
during the first 2--3 Myr of the evolution and the radiogenic heating leads
first to the melting of the ice to form a clay body surrounded by a surface ice
layer.  Water released from the melting of primitive MetSilWat material is
pushed to the surface where it is cooled by the vicinity to the surface. This is
followed by interior heating to above the melting temperature of silicates and
the separation of the interior into an early core and mantle.  However, the
draining of the radiogenic energy prohibits further differentiation and instead
the accreted layers are heated only enough to forming clay and adding extensive
water to the surface ice layer. The effective accretion temperature finally
becomes high enough to trigger a run-away greenhouse effect in the coupled
ocean-atmosphere system after 3.4 Myr, heating the surface to above the melting
temperature of the silicates. This leads to separation of metal melt from
silicate melt and the sedimentation of the metal to form the core\footnote{ Our
model instantaneously moves released metal down to the base of the magma ocean.
\cite{Rubie+etal2003} demonstrated that liquid metal dispersed in magma forms
droplets of 1 cm in diameter that sediment at a speed of approximately 0.5 m/s.
This gives a transport time from magma ocean to core of approximately 0.1 yr for
an Earth-sized planet, which is much faster than the supply of metal by pebble
accretion. Hence our assumption of instantaneous transport of metal to the core
is valid.}. The released gravitational energy by early core formation powers
additional melting of the remaining solid mantle and the full interior
separation of the entire planet into a core and mantle.

\subsection{Energetics of differentiation}

The gravitational potential energy difference of a blob of metal with mass
$m_{\rm met}$ falling from the surface of a protoplanet with mass $M$ and radius
$R$ to the core-mantle boundary at radius $R_{\rm c}$ is
\begin{equation}
   \Delta E = \frac{G M m_{\rm met}}{R} \left[ \frac{3}{2} - \frac{1}{2}
   \left(\frac{R_{\rm c}}{R} \right)^2 \right] -\frac{G M m_{\rm met}}{R} \, .
\end{equation}
This energy release must be equated with the temperature increase of both the
metal (of mass $m_{\rm met}$) and of the silicate component co-accreted with the
metal (of mass $m_{\rm sil}$),
\begin{equation}
  \Delta E = c_{\rm met} m_{\rm met} \Delta T + c_{\rm sil} m_{\rm sil} \Delta T
  \, .
\end{equation}
In the limit $R_{\rm c} \ll R$ we can write the temperature increase as
\begin{equation}
  \Delta T = \frac{G M}{2 R (c_{\rm met}+c_{\rm sil} m_{\rm sil}/m_{\rm met})} =
  1.1 \times 10^4\, {\rm K} \left( \frac{M}{M_{\rm E}} \right)^{2/3} \left(
  \frac{\rho}{\rho_{\rm E}} \right)^{1/3} \, .
\end{equation}
Here we normalized to the compressed density of Earth, $\rho_{\rm E}$, and used
$c_{\rm met}=800\,{\rm J\,kg^{-1}\,K^{-1}}$, $c_{\rm sil}=1200\,{\rm
J\,kg^{-1}\,K^{-1}}$, and $m_{\rm sil}/m_{\rm met} = 0.62/0.38$ (this includes
the full S inventory of the metal, see Paper I). The temperature increase is
reduced to $\Delta T \approx 2.4 \times 10^3\,{\rm K}$ at $M = 0.1 M_{\rm E}$
and $\Delta T \approx 500\,{\rm K}$ at $M = 0.01 M_{\rm E}$. The initial fall
from infinity to the surface releases twice the energy per mass compared to the
subsequent descent of the metal to the core. Since this fall affects the
entire mass budget, the accretion to the surface has the potential to heat
approximately four times more than the metal descent \citep{Rubie+etal2015b}.
Part of this energy is nevertheless carried away through the atmosphere and
envelope. The initial differentiation in our model is mainly driven by the
penetration into the solid mantle of the metal that melted at the surface. As
the planet grows, the trapped accretion energy suffices to melt metal already at
its arrival at the surface and the descent to the core only serves to heat the
molten metal additionally.

\subsection{Envelope mass}

We show the mass of the gas envelope attracted from the protoplanetary disc in
Figure \ref{f:Menv_t}. Earth grows massive enough to attract a significant
envelope after 3 Myr of evolution after reaching approximately 0.01 $M_{\rm E}$.
The mass then increases rapidly as the Bondi radius $R_{\rm B} = G M / c_{\rm
s}^2$ expands with increasing mass, peaking at over $10^{19}\,{\rm kg}$.  The
gas is nevertheless removed very rapidly by XUV irradiation from the young star
after the dissipation of the protoplanetary disc. In contrast, the small mass of
Mars prevents accretion of any significant gas envelope.

\subsection{Core mass and surface temperature}

In Figure \ref{f:mass_t} we show the masses of the planet, core, magma ocean and
crystallized mantle, as well as the surface temperature and envelope
temperature, as a function of time for our Mars and Earth analogues. 
Accretion of new layers of primitive, ice-rich material in the first million
years leads to rapid melting, outgassing and loss of water. This is visible as
intermittent temperature spikes that are caused by our discrete mass accretion
scheme (see Paper I).

The core mass increases with time during the first 3 Myr, but the depletion of
the $^{26}$Al reservoir then prevents further differentiation. The outgassed
atmosphere nevertheless undergoes a run-away greenhouse effect as the effective
accretion temperature reaches approximately 300 K. This leads to a rapid melting
of the mantle and formation of a core containing the bulk iron budget of the
planet. The magma ocean crystallizes rapidly after the end of the accretion
phase. This leads to the formation of an early mantle with a melting degree of
$\varPhi \approx 0.4$ where the thermal conductivity falls dramatically below
the fully molten value. The rapid magma ocean crystallization of Mars within a
million years after the end of accretion agrees well with age measurements of
zircons from martian meteorites \citep{Bouvier+etal2018,Zhu+etal2022}.

\begin{figure}
  \includegraphics[width=0.9\linewidth]{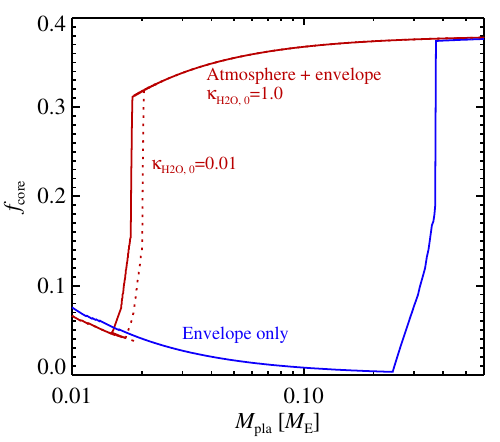}
  \caption{A comparison of the core mass fraction as a function of the accreted
  mass between the standard model, where we include both the outgassed
  atmosphere and the hydrostatic envelope connected to the protoplanetary disc,
  and an additional calculation where we do not include the blanketing effect of
  the outgassed atmosphere. Neglecting the atmosphere raises the critical planet
  mass for differentiation by a factor of 10 from approximately $0.02$ $M_{\rm
  E}$ to $0.3$ $M_{\rm E}$. We also show the evolution of the core mass fraction
  for a water opacity level 100 times lower than our nominal value. The rapid
  increase of the accretion luminosity with mass means that even a substantial
  lowering of the opacity translates to only a small increase in the critical
  mass that triggers the run-away greenhouse effect and commences
  differentiation.}
  \label{f:core_fraction_t}
\end{figure}

To illustrate the importance of the outgassed atmosphere in facilitating
differentiation by thermal blanketing, we show in Figure \ref{f:core_fraction_t}
the core mass fraction as a function of planetary mass for our Earth model. We
compare in the figure the model including both the atmosphere and the envelope
with a calculation where we include only the hydrostatic gas envelope connected
to the protoplanetary disc. Ingassing from contact between such an envelope and
the magma ocean has been proposed to explain the water D/H ratio of Earth
\citep{Wu+etal2018} as well as the presence of noble gases of solar composition
in the deep mantle \citep{WilliamsMukhopadhyay2019}. In the absence of a
blanketing atmosphere, the protoplanet must grow to approximately 0.3 $M_{\rm
E}$ before the surface temperature reaches a high enough energy to melt the
mantle and separate metal from silicates. This contrasts with the ten times
lower critical differentiation mass of approximately $0.02$ $M_{\rm E}$ when the
outgasssed atmosphere is included in the calculations. Therefore a small planet
like Mars could only differentiate through the blanketing effect of the
outgassed atmosphere, while Earth and Venus are large enough to differentiate
from the heat trapped by the hydrostatic envelope alone.

\subsection{Atmosphere structure}

\begin{figure}
  \includegraphics[width=0.9\linewidth]{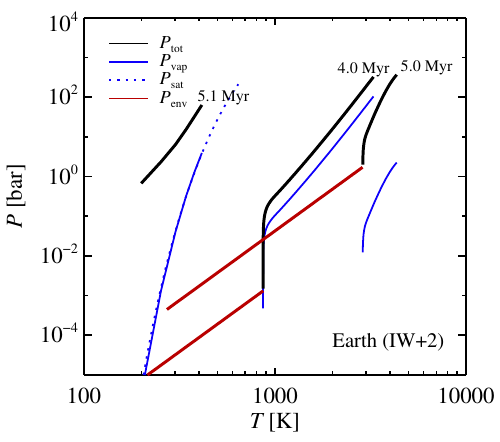}
  \caption{The pressure as a function of temperature for our Earth analogue
  at three selected times. Before the dissipation of the protoplanetary disc, at
  $t=4.0$ Myr and $t=5.0$ Myr, the atmosphere connects to the envelope at a
  pressure and temperature that increases with time as the hydrostatic envelope
  becomes more massive. The removal of the gas envelope after disc dissipation
  at $t=5.0$ Myr leads to a much lower pressure and temperature at the
  photosphere of the isolated planet.  The partial pressure of water vapour
  (thin blue line) follows the shape of the atmospheric pressure for high
  temperatures (translated down in pressure to reflect the water vapour mixing
  ratio) and the saturated pressure (dotted blue line) for low temperatures.}
  \label{f:atmosphere_structure_Earth}
\end{figure}

The structure of the planetary atmosphere and envelope goes through many
phases during the accretion. We show the pressure-temperature relation in Figure
\ref{f:atmosphere_structure_Earth} for four illustrative times. During the
accretion phase, the hydrostatic gas envelope exerts a pressure on the
atmosphere that increases with the mass of the protoplanet and hence with time.
The temperature at the base of the envelope is larger than the critical
temperature of water, hence water is always in gaseous form in the later stages
of the accretion. At $t=5.0$ Myr the pressure reaches $10^4$ bar and the
temperature $6,000$ K at the surface. The atmospheric pressure is here dominated
by SiO vapour while the water vapour constitutes only a small fraction of the
pressure. After the dissipation of the protoplanetary disc, the envelope
pressure is released and the conditions at the atmosphere's photosphere, where
radiation can leave freely to space, drop down to the effective temperature of
the planet under solar irradiation. The first water cloud layers form here as
the partial pressure of water crosses over the saturated vapour pressure line.

\begin{figure}
  \begin{center}
    \includegraphics[width=0.9\linewidth]{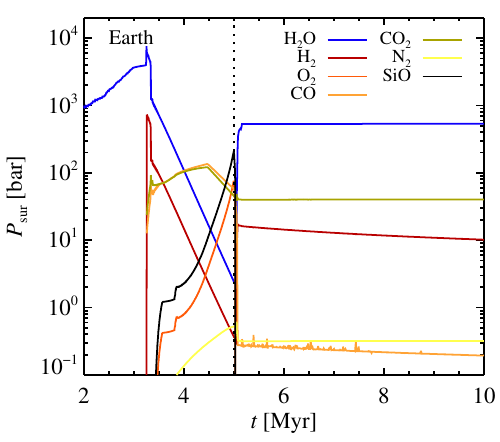}
  \end{center}
  \caption{The partial atmospheric surface pressure for all the considered
  atmospheric components for our Earth analogue. The water pressure includes
  here both the vapour atmosphere and the underlying water ocean. The initial
  spike in water pressure is due to the run-away greenhouse effect caused by the
  accretion energy. This initial outgassed atmosphere is thereafter dissolved in
  the magma ocean and replaced by an atmosphere with important contributions
  from SiO (from evaporation of the magma ocean) and O$_2$ (from the
  iron-w\"ustite buffer in the magma ocean). The cessation of the accretion
  energy after 5 Myr leads to a rapid decline of the SiO/O$_2$ atmosphere,
  followed by an outgassing of the final atmosphere dominated by CO$_2$ and
  H$_2$O.}
  \label{f:Psur_t}
\end{figure}

The partial pressures of the considered atmospheric components at the
surface are shown in Figure \ref{f:Psur_t}. The onset of the run-away greenhouse
effect in the water component is visible after 3 Myr as a spike in the partial
pressure of water. The subsequent emergence of the magma ocean leads to
efficient reburial of the H$_2$O in the magma followed by a slow increase of
both SiO and O$_2$ pressure in the atmosphere. The increase in O$_2$ pressure is
due to the increase of the iron-w\"ustite buffer in the top of the magma ocean
with increasing temperature. The SiO/O$_2$ atmosphere `condenses' back onto
the magma ocean after the stop of the accretion phase.  This is followed by the
outgassing of the primordial atmosphere of the planet, dominated by CO$_2$, CO
and H$_2$O as set by the assumption of oxidation stage fO2 = IW + 2 in the
mantle. The gradual reduction of the magma volume leads to outgassing first of
N- and C-bearing species, while the high solubility of water means that this
molecule is outgassed mainly towards the end of the magma ocean crystallization
phase \citep{Bower+etal2022}.

\subsection{Interior temperature and magma ocean crystallization}

\begin{figure*}
  \begin{center}
    \includegraphics[width=0.9\linewidth]{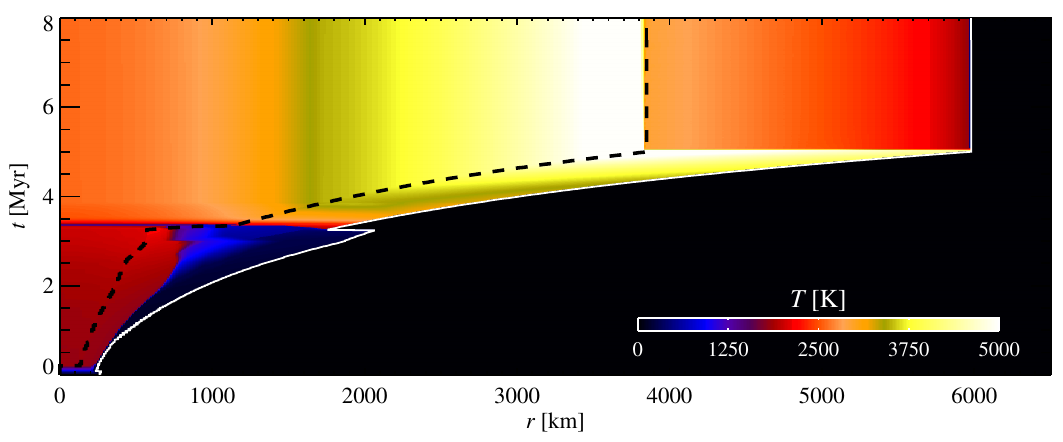}
  \end{center}
  \caption{Temperature map of our Earth analogue as a function of distance form
  the centre and time. The black dashed line indicates the size of the core,
  while the planetary radius is indicated with a thin white line. The accretion
  of cold material, unable to heat to differentiation by $^{26}$Al decay, is
  visible as a blue clump between 2 Myr and 3.5 Myr. The planetary radius falls
  significantly as the water ocean is lost to the run-away greenhouse effect
  after 3.4 Myr. The termination of accretion after 5 Myr is followed by a slow
  cooling first and then by rapid cooling when the gas envelope escapes after 7
  Myr. This leads to a decrease of the mantle temperature towards the solidus
  value where the heat conductivity due to convective heat transport drops
  dramatically. The core remains in a hot state after mantle cooling, since the
  hot outer core stabilizes the liquid metal against convective heat transport.}
  \label{f:TT_r_t}
\end{figure*}

The interior temperature evolution of our Earth analogue is shown in Figure
\ref{f:TT_r_t}. The space-time plot shows the instantaneous value of the
temperature at times up to 10 Myr. The early core formation is visible as a
region of 2,000 K within the growing planet. This is followed by cold accretion
of material that only reaches the melting temperature of water but not the
differentiation temperature. The radius of the planet falls as the run-away
greenhouse effect sets in after 3.4 Myr and transfers all the water from the
surface ocean to the supercritical steam atmosphere. This is followed by the
heating of the planet from the surface and downwards, an effect which is
exacerbated by the sedimentation of iron droplets that release gravitational
binding energy. After the accretion terminates, the mantle temperature falls
rapidly to near the solidus temperature of the silicates.

The mode of closing of the magma ocean plays an important role in distributing
volatiles between the mantle and the atmosphere \citep{Elkins-Tanton2008}. The
magma ocean is most prone to solidify from the bottom, due to higher solidus
temperature at higher pressure and to the foundering and remelting of any
emerging crust into the liquid magma below \citep{WeissElkins-Tanton2013}. We
hence assume that a magma ocean extending up to at least 70\% of the planetary
radius is in contact with the atmosphere irrespective of whether a lid forms in
the numerical simulations. We close the contact between the mantle and the core
after a solid layer forms at the bottom of the mantle with a thickness of at
least 10\% of the planetary radius.

\begin{figure}
  \begin{center}
    \includegraphics[width=0.9\linewidth]{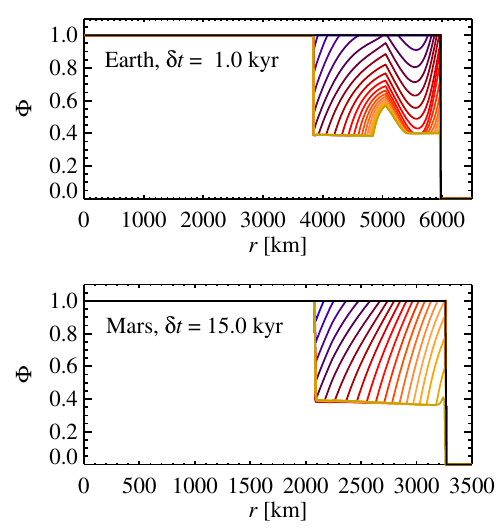}
  \end{center}
  \caption{The interior melting degree $\varPhi$ of our Earth and Mars analogues
  in steps of 1 kyr (Earth) and 15 kyr (Mars) after first reaching the magma
  ocean liquidus after termination of accretion. Earth experiences rapid
  crystallization and a solidification that propagates first from the top of the
  magma ocean and down and later from the bottom up. A small region is left in
  the middle mantle with a melting degree that slowly approaches $\varPhi = 0.4$
  below which the magma viscosity increases strongly. The Mars analogue
  crystallizes its magma ocean from the bottom up and more slowly, due to
  thermal blankering by its thick H$_2$-rich atmosphere that slows the cooling
  of the magma ocean.}
  \label{f:PPhi_r}
\end{figure}

We show in Figure \ref{f:PPhi_r} the evolution of the melting degree of the core
and the mantle after the gas envelope has been removed and the liquidus
temperature is first reached in the magma ocean. The magma ocean of our Earth
analogue crystallizes initially from surface down and later from the core and
up. This traps a small melt region in the middle of the mantle that slowly
approaches the melting degree $\varPhi \approx 0.4$ where mantle viscosity
increases dramatically (see Paper I). Our Mars analogue, in contrast,
crystallizes its magma ocean from the bottom up.

\subsection{Evolution of the water reservoirs}

\begin{figure}
  \begin{center}
    \includegraphics[width=0.9\linewidth]{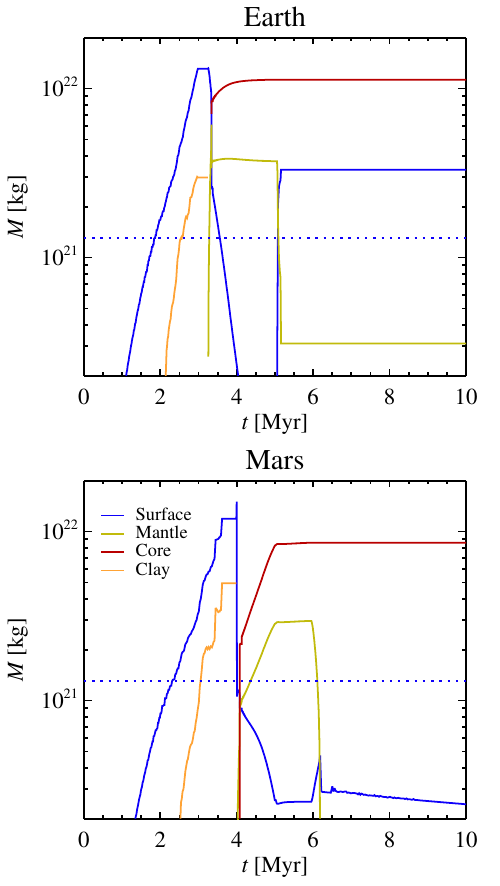}
  \end{center}
  \caption{The mass of water in different reservoirs as a function of time (top:
  Earth, bottom: Mars). Before the onset of the run-away greenhouse effect, most
  of the water resides in the massive surface layer and in the clay mantle
  below. The majority of the water is dissolved into the newly formed magma
  ocean and core when the planets differentiate by the accretion energy. The
  crystallization of the magma ocean after 7 Myr leads to the release of
  approximately two modern Earth oceans of water on Earth, while the reduced
  mantle of Mars leads to outgassing of far less water.}
  \label{f:Mwater_t}
\end{figure}

We will analyse the volatile (H, C, O, and N) contents of core, mantle, and
atmosphere, as well as the early atmospheric escape, in more detail in Paper
III. Here we describe only on the fate of water within the growing planets. Our
planets accrete approximately $2 \times 10^{22}$ kg of water from icy pebbles
\citep{Johansen+etal2021}, which corresponds to 15 Earth oceans. This amount is
independent on the final mass of the planet, so Mars and Earth accrete the same
amount of water despite the much lower mass of Mars. We show in Figure
\ref{f:Mwater_t} the mass of the water in the different reservoirs. Before the
run-away greenhouse effect, most of the water resides in the surface ocean and
the clay mantle. This water is lost to the atmosphere as supercritical steam
after the run-away greenhouse effect sets in. However, the formation of the
global magma ocean opens a new reservoir for water -- and most of the water
dissolves in the magma ocean from where a large fraction also enters the core.
The core of Earth can hold much more water than the smaller core of Mars; hence
95\% of the water resides in the end in the core of Earth, while Mars' core
holds a factor two less water. The crystallization of the magma ocean after the
accretion terminates pushes the incompatible water back into the atmosphere. We
assume here that the upper mantle of Earth has an oxygen fugacity of IW+2
\citep{Armstrong+etal2019,Sossi+etal2020,Deng+etal2020}, while Mars is much more
reduced with IW-2 because its magma ocean does not reach depths where
Fe$^{3+}$ is produced in reactions between less-oxidized Fe$^{2+}$. Hence Earth
outgasses approximately two modern oceans of water, while Mars outgasses mainly
hydrogen that is lost by hydrodynamic escape. The composition and escape of the
outgassed atmosphere are discussed in details in Paper III.

\begin{figure}
  \begin{center}
    \includegraphics[width=0.9\linewidth]{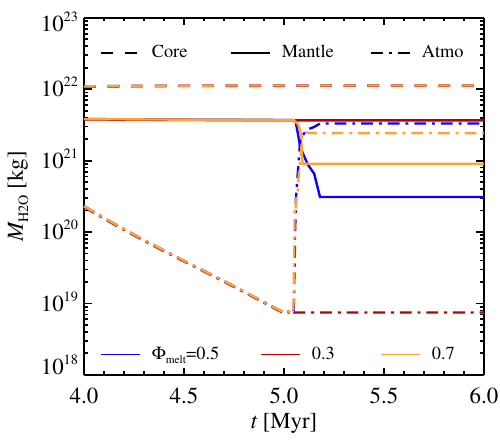}
  \end{center}
  \caption{The mass of water in the core, mantle and atmosphere reservoirs
  for three different choices of the threshold melt fraction $\varPhi_{\rm
  melt}$ above which the magma ocean is considered liquid and can hence exchange
  volatiles with the core and the atmosphere. The core reservoir is relatively
  unaffected by this choice and holds in all cases the dominant water
  mass. Lowering the threshold fraction to 0.3 means that the magma ocean does
  not formally crystallize and hence the mantle reservoir cannot outgas
  efficiently to the atmosphere. Increasing instead to $\varPhi_{\rm melt}=0.7$
  causes less outgassing than the nominal choice of $\varPhi_{\rm melt}=0.5$,
  since the volatile exchange with the atmosphere ceases at a higher magma mass
  that can thus store more volatiles. This reduces the atmospheric reservoir of
  water by a factor of approximately two.}
  \label{f:core_mantle_reservoirs_melt}
\end{figure}

We have assumed that the partitioning between core, mantle and atmosphere
takes place only when the magma ocean melt fraction is higher than a threshold
value of $\varPhi = \varPhi_{\rm melt}=0.5$ (see Section \ref{s:dissolution}).
This choice is motivated by a transition from liquid behaviour to solid
behaviour around a critical melt fraction of $\varPhi=0.4$ \citep[][see
discussion in paper I]{Monteux+etal2016}. To test the influence of this choice
on the partitioning of water, we show in Figure
\ref{f:core_mantle_reservoirs_melt} the mass of water stored in the core, mantle
and atmosphere for the nominal choice of $\varPhi_{\rm melt}=0.5$ compared to
$\varPhi_{\rm melt}=0.3$ and $\varPhi_{\rm melt}=0.7$. When the threshold melt
fraction $\varPhi_{\rm melt}$ is lowered, the magma ocean formally can not
crystallize after accretion, since the collapse of the heat conductivity at
$\varPhi \approx 0.4$ prevents further, rapid cooling. Increasing instead the
threshold melt fraction to $\varPhi_{\rm melt}=0.7$ leads to a surge in the
mantle reservoir of water, as the water concentration ``freezes'' in at a higher
magma mass that can hence hold more water. The atmospheric reservoir is
therefore reduced by a factor of approximately two compared to the nominal
choice of melt fraction threshold.

\subsection{Conditions for core-mantle separation}

The concentrations of moderately and weakly siderophile elements in the mantle
of Earth can be used to constrain the pressure at which metal and silicate melts
equilibrated their elemental concentrations. \cite{Mann+etal2009} concluded that
the differentiation pressure must have exceeded 30--60 GPa to match their
experimental measurements of the metal-liquid partition coefficients of a range
of siderophile elements. \cite{Badro+etal2015} obtained similar constraints
based on the contents of Si and O in the terrestrial core. The pressure at the
core-mantle boundary of our Earth analogue with a mass of $0.6\,M_{\rm E}$ is 60
GPa, in good agreement with the conclusion of \cite{Mann+etal2009} and
\cite{Badro+etal2015}. However, the core-mantle boundary pressure after the
moon-forming giant impact must have been similar to the modern value (136 GPa)
and this raises the question to what degree the metal of the core of Theia
equilibrated with the mantle material of Earth and Theia during the collision.
We discuss mantle-core equilibration of W isotopes during the moon-forming
impact in the next section.

The temporally evolving oxidation state of the magma ocean also affects the
partitioning of elements such as Si and O between liquid metal and silicates.
\cite{Badro+etal2015} presented mantle-core partitioning results for a range of
temporal profiles of the FeO fraction in the magma. They found the best matches
to the abundance of siderophile elements in the mantle and the density of the
core (lowered due to incorporation of lighter elements) for FeO mantle fractions
initially around 20--25\% and falling towards 6\% with time, with similar
pressure constraints as \cite{Mann+etal2009}. This increasingly reduced mantle
is in good agreement with the pebble accretion model where most of the FeO is
produced by water flow during the earliest accretion and subsequently thinned
out by accretion of dry pebbles with a high content of metallic iron (see Paper
I)\footnote{We note that the increase in the fraction of Fe$^{3+}$ relative to
Fe$^{2+}$ caused by reactions between less-oxidized Fe$^{2+}$ in the deep magma
oceans of Earth and Venus increases the oxygen fugacity near the surface, while
the bulk magma ocean is overall is still reduced \citep{Armstrong+etal2019}.}.
This model contrasts with the prevailing view that Earth became increasingly
oxidized during its accretion \citep{WadeWood2005,Rubie+etal2015a}.  For the
weakly and moderately siderophile elements in the mantle of Mars, modelling is
generally consistent with differentiation under reducing conditions \citep[with
oxygen fugacity around ${\rm IW}-1.5$][]{RighterChabot2011,RaivanWestrenen2013}.
This corresponds roughly to the inferred FeO mass fraction in the mantle of Mars
of 18\% \citep{RobinsonTaylor2001} which we reproduced in the pebble accretion
model in Paper I.

\section{Hf-W model fits}

Hf is a lithophile (rock-loving) element with an unstable isotope $^{182}$Hf
that decays to $^{182}$W with a half-life of approximately 8.9 Myr. Since W is
moderately siderophile (iron-loving), the evolution of the abundances of Hf and
W in the mantle and core of our Earth analogue can be linked to measurements of
the abundances of these elements in Earth's mantle and in chondrites, in order
to infer the consistency of our model of rapid terrestrial planet formation
against those measurements. We follow \cite{Yin+etal2002} and define the excess
of $^{182}$W relative to $^{183}$W in the mantle of our model planets as
\begin{equation}
  \epsilon_{\rm W} = \left[ \frac{(^{182}{\rm W}/^{183}{\rm W})_{\rm
  model}}{(^{182}{\rm W}/^{183}{\rm W})_{\rm Earth}} -1 \right] \times 10^4 \,
  .
\end{equation}
Here $(^{182}{\rm W}/^{183}{\rm W})_{\rm Earth} \approx 1.85130$ is the measured
ratio in Earth's mantle, which by this definition has $\epsilon_{\rm
W}=0.0$. The initial value at the birth of the Solar System,
before substantial $^{182}$Hf decay, was
$\epsilon_{\rm W,SS}=-3.5$ and the modern value for undifferentiated chondritic
material is $\epsilon_{\rm W,CHUR}=-2.0$. The low value of $\epsilon_{\rm W}$
for Earth, compared to, for example, Vesta with $\epsilon_{\rm W}=17$
\citep{Jacobsen2005}, has been used to argue for a protracted Earth accretion
taking at least 34 Myr \citep{KleineWalker2017}, in agreement with traditional
planetesimal accretion models \citep{Raymond+etal2004}, or for rapid accretion
within 10 Myr \citep{Yin+etal2002,YuJacobsen2011}, followed by equilibration in
the moon-forming impact. This giant impact could have significantly lowered the
$\epsilon_{\rm W}$ value by interaction of Earth's mantle with material from
the $^{182}$W-poor core of the impacting planetary body (see discussion below).
\begin{figure*}
  \begin{center}
    \includegraphics[width=0.7\linewidth]{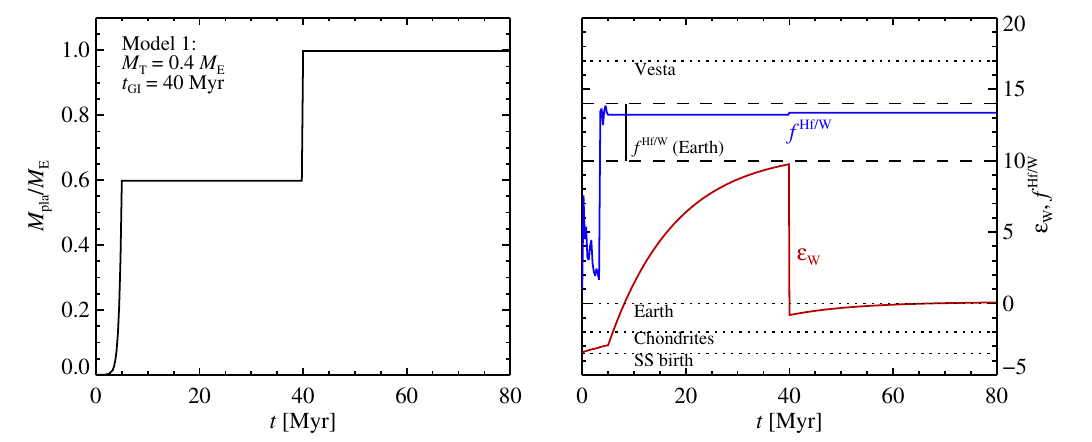}
    \includegraphics[width=0.7\linewidth]{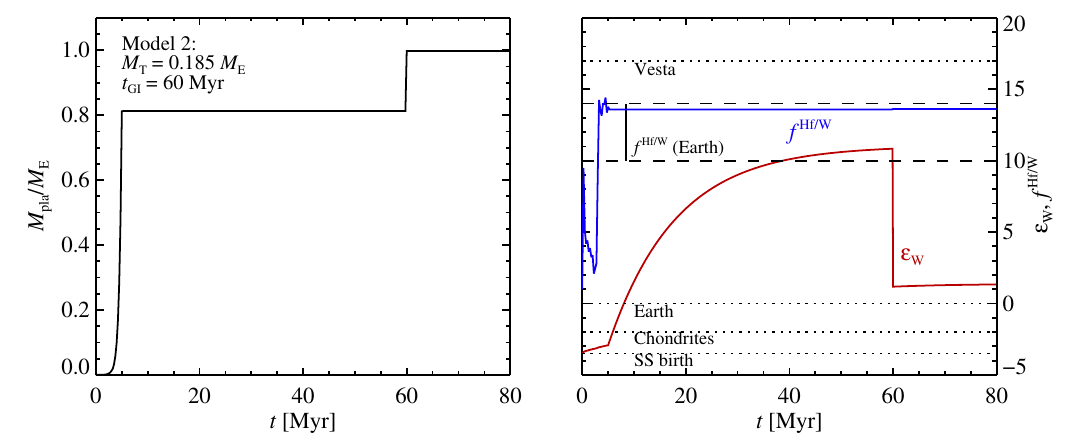}
    \includegraphics[width=0.7\linewidth]{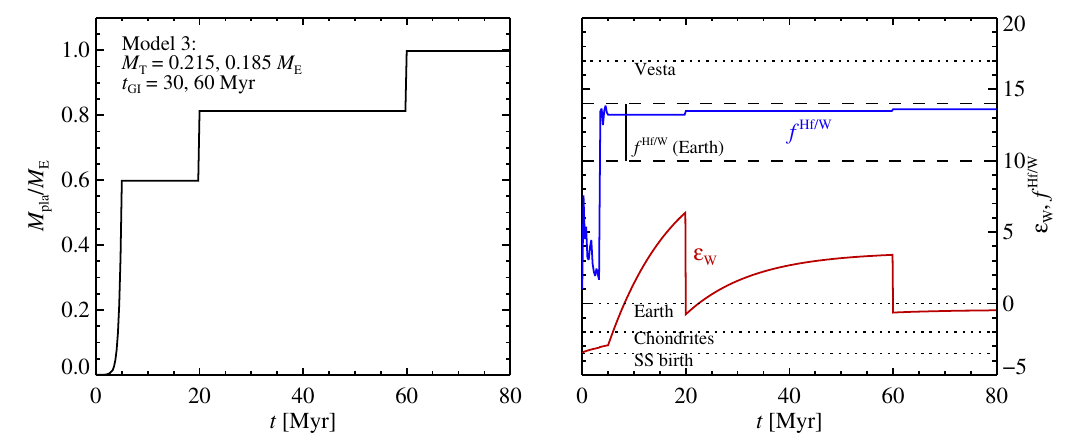}
    \includegraphics[width=0.7\linewidth]{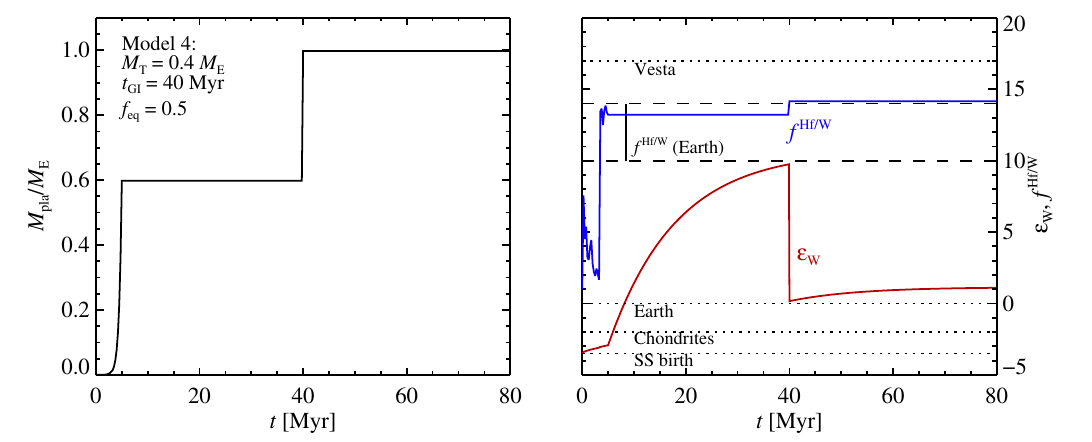}
  \end{center}
  \caption{Comparisons of four models of Earth formation by rapid pebble
  accretion to the Hf-W isotope data for Earth. The top panels (Model 1) shows
  results for our nominal impactor mass of $0.4\,M_{\rm E}$ colliding with
  proto-Earth after 40 Myr. The excess of $^{182}$W in the mantle builds up a to
  a value of nearly 10 before the interaction of the mantle material with the
  impacting Theia core delivers enough $^{183}$W to bring the $\epsilon_{\rm W}$
  value down towards the measured data. The second panel (Model 2) show instead
  the effect of lowering the impactor to $0.185\,M_{\rm E}$ (we use here our
  Venus analogue for the proto-Earth model). We set the impact to occur after 60
  Myr to avoid subsequent build up of $^{182}$W in the mantle, but the impactor
  mass is too low to bring the model to agree with the measured $\epsilon_{\rm
  W}$ of Earth. The canonical impactor model is reconciled with the Hf-W system
  in the third row (Model 3) where we include two giant impacts: an early impact
  at 20 Myr removes the initial $^{182}$W excess and the second (moon-forming)
  giant impact brings the subsequent mantle ingrowth of $^{182}$W down to the
  modern terrestrial value. Finally, the lower panels (Model 4) show the
  effect of lowering the equilibration efficiency to $f_{\rm eq}=0.5$. The
  $\epsilon_{\rm W}$ value ends above the terrestrial value in this case.} 
  \label{f:Hf_W_t}
\end{figure*}
\begin{figure*}
  \begin{center}
    \includegraphics[width=0.9\linewidth]{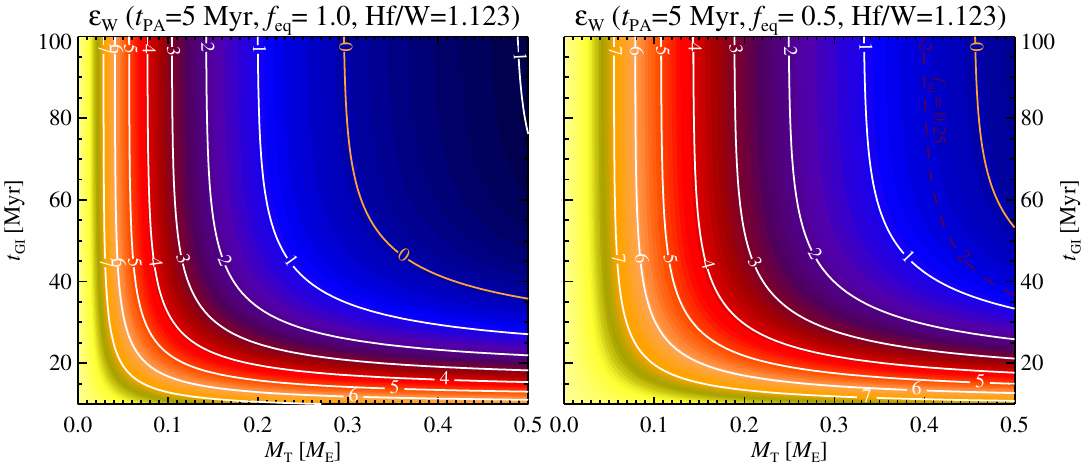}
  \end{center}
  \caption{Contour plot of the $^{182}$W excess of Earth's mantle after 200 Myr
  of evolution, as a function of the impactor mass ($M_{\rm T}$) and the time of
  the giant impact ($t_{\rm GI}$). The left plot shows results for complete
  equilibration between the core of Theia and proto-Earth, while the right plot
  shows results for 50\% equilibration (i.e., 50\% of the core of Theia merges
  directly with Earth's core). The pebble accretion time-scale has been set to
  $t_{\rm PA}=5$ Myr in both cases. The $\epsilon_{\rm W}=0$ value of Earth is
  consistent with the pebble accretion model for an impactor mass between
  $0.3\,M_{\rm E}$ and $0.5\,M_{\rm E}$ and a fully equilibrating impact that
  occurs at least 35 Myr after the formation of the Sun. Lowering the
  equilibration to 50\% requires an impact between two nearly equal-mass bodies
  to occur after at least 50 Myr.  The so-called canonical model, with an
  impactor of 10\% Earth mass \citep{CanupAsphaug2001}, is excluded in both
  cases which instead favour an impactor with a similar mass as proto-Earth
  \citep{Canup2012}. A very low equilibration factor of $f_{\rm eq}=0.25$
  yields $\epsilon_{\rm W}$ values larger than 1.5 for all impactor masses; the
  $\epsilon_{\rm W}=2.0$ contour for $f_{\rm eq}=0.25$ is indicated with a black
  dashed line.}
  \label{f:Hf-W_simple}
\end{figure*}

\subsection{Modelling the Hf-W system}

We postprocess the evolution of the Hf-W system of our Earth analogue by
dividing the accretion into several time-steps and calculating for each
time-step how the accreted W distributes between the core and the mantle and how
the abundance of $^{182}$W in the mantle increases by the decay of $^{182}$Hf.
We assume that an accreted mass $\delta M$ consists of a fraction $f_{\rm
Sil}=68\%$ silicates and a fraction $f_{\rm Met}=32\%$ metal and equilibrates its
contents of W with a whole-mantle magma ocean. The accreted core mass is $\delta
M_{\rm core} = f_{\rm Met} \delta M$ and the amount of W accreted into the core is
\begin{equation}
  \delta M_{\rm W}^{\rm (core)} = C_{\rm W}^{\rm (Met)} \delta M_{\rm core} \, .
\end{equation}
Here $C_{\rm W}^{\rm (Met)}$ is the equilibrium concentration of W in the metal
melt. Knowledge of the total amount of accreted W and its partition coefficient
$D = C_{\rm W}^{\rm (Met)}/C_{\rm W}^{\rm (Sil)}$ yields the connection
\begin{equation}
  M_{\rm W}^{\rm (Sil)} + \delta M_{\rm W} = C_{\rm W}^{\rm (Sil)} (M_{\rm Sil}
  + \delta M_{\rm Sil}) + C_{\rm W}^{\rm (Met)} \delta M_{\rm core} \, .
\end{equation}
Here $M_{\rm W}^{\rm (Sil)}$ is the existing W in the mantle, $\delta M_{\rm W}$
is the accreted W, $M_{\rm Sil}$ is the mass of the mantle and $\delta M_{\rm
Sil}$ is the accreted silicates. Replacing $C_{\rm W}^{(\rm Sil)} = C_{\rm
W}^{\rm (Met)} / D = \delta M_{\rm W}^{\rm (core)} / (D \delta M_{\rm core})$
yields the W addition to the core
\begin{equation}
  \delta M_{\rm W}^{\rm (core)} = \frac{M_{\rm W}^{\rm (Sil)}+\delta M_{\rm
  W}}{1+(M_{\rm Sil}+\delta M_{\rm Sil})/(D \delta M_{\rm core})} \, .
\end{equation}
This expression covers both single accretion event where the whole impactor
(core and mantle) equilibrates with the whole mantle of the planet before the
metal component sinks to the core, as well as continuous pebble accretion where
$\delta M_{\rm Sil} \ll M_{\rm Sil}$ due to the time-stepping. In that limit the
accretion expression tends towards
\begin{equation}
  \delta M_{\rm W}^{\rm (core)} \rightarrow  D \frac{M_{\rm W}^{\rm
  (Sil)}}{M_{\rm Sil}} \delta M_{\rm core} \, .
\end{equation}

The partition coefficient $D$ of W between metal melt and silicate melt is
relatively uncertain and depends on both equilibration pressure, temperature and
the presence of other elements such as O, S and C in the melt
\citep{WalterCottrell2013,Jennings+etal2021}. We therefore for simplicity take
$D = 25$; this gives a good match to the Hf/W overabundance of Earth relative to
the chondritic value, $f^{\rm Hf/W} = 12 \pm 2$ \citep{Jacobsen2005}.
\begin{figure*}
  \begin{center}
    \includegraphics[width=0.9\linewidth]{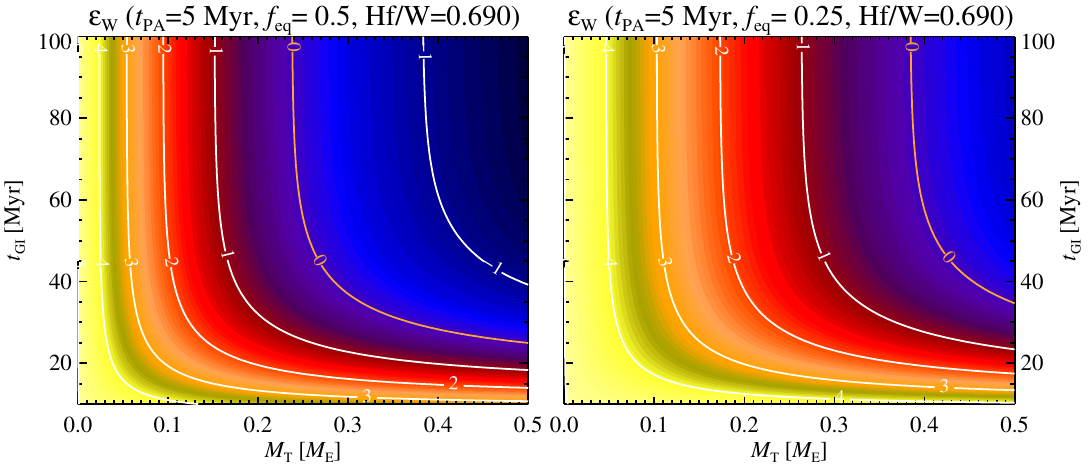}
  \end{center}
  \caption{The final $^{182}$W excess as a function of impactor mass and
  impact time, for an initially reduced value of Hf/W=0.69 that may represent
  the inner Solar System composition \citep{NimmoKleine2007}. The lower amount
  of $^{182}$Hf makes it easier for the impactor core material to transport
  excess $^{182}$W to the combined post-giant-impact core. Hence the pebble
  accretion time-scale of 5 million years is here consistent with core-mantle
  equilibration factors as low as $f_{\rm eq}=0.5$ (left plot) and $f_{\rm
  eq}=0.25$ (right plot).}
  \label{f:Hf-W_simple_lowHfW}
\end{figure*}

We show the results of our Hf-W modelling in Figure \ref{f:Hf_W_t}. The results
depend strongly on the choice of impactor mass and the degree of
equilibration between the impactor core and the mantle material. To bracket the
realm of possible matches to the Hf-W data, we consider the mass of Theia to
have been either $0.4\,M_{\rm E}$ (our nominal case) or $0.185\,M_{\rm E}$. In
the latter case we use our Venus analogue as a model of proto-Earth before the
collision. We can match the measured value of $\epsilon_{\rm W}$ for Earth to
rapid core formation on Earth within 5 Myr, which drives $\epsilon_{\rm W}$ up
to 9--10, followed by a giant impact happening at the earliest after $t=35$ Myr
where the impactor equilibrates fully with Earth and hence drives down
$\epsilon_{\rm W}$ to the modern value. Lowering the impactor mass to
$0.185\,M_{\rm E}$ precludes any match to the data, since the impactor does not
contain enough W in its core to equilibrate Earth's value of $\epsilon_{\rm
W}$ down to 0.

\subsection{Testing impactor mass and timing}

The consistency of the rapid pebble accretion model for Earth formation with the
Hf-W system clearly hinges on the exchange of W isotopes between the core of the
moon-forming impactor and the mantle of Earth \citep{Rudge+etal2010}. In
Figure \ref{f:Hf-W_simple} we map the final value of the $^{182}$W excess as a
function of the mass of the impactor and the time of the giant impact. We
used here a simple growth model where the pebble accretion rate scales with
planetary mass as $\dot{M} \propto M^{2/3}$, approximating growth in the 2-D
Hill regime of pebble accretion \citep{LambrechtsJohansen2012}. This gives the
growth function $M(t) = ( 3 K t)^{1/3}$, where $K$ is a growth constant that is
fixed by requiring that the planet reaches its pre-impact mass $M_1$ after a
time of $t_1=5\,{\rm Myr}$.  Planetesimal-driven models typically use an
exponentially decaying mass curve \citep{Kleine+etal2009}; we have verified that
the specific choice of growth curve affects the final $\epsilon_{\rm W}$ value
by less than 0.1.

Figure \ref{f:Hf-W_simple} shows results for core-mantle equilibration of 100\%
and 50\% in the giant impact. In the latter case, half of the core of Theia
merges with the core of Earth without exchanging any isotopes with the mantle.
For full equilibration, the pebble accretion model is consistent with an
impactor mass between $0.3\,M_{\rm E}$ and $0.5\,M_{\rm E}$ happening at least
35 Myr after the formation of the Sun. For the lower end of this range, the
giant impact time can be very long, since the impact itself drives
$\epsilon_{\rm W}$ to near zero. A young age of the Moon is also consistent with
Pb-Pb dating yielding moon formation $\sim$150 Myr after the formation of the
Solar System \citep{ConnellyBizzarro2016}.  Lowering the equilibration to 50\%,
consistency of the pebble accretion model with the Hf-W system requires two
nearly equal-mass planets to collide after at least 50 Myr.

Figure \ref{f:Hf-W_simple} used the initial ratio Hf/W$=1.123$ measured for
carbonaceous chondrites \citep{Yin+etal2002}. Ordinary chondrites, representing
the composition of the inner Solar System, display a range of lower and higher
values. We follow \cite{NimmoKleine2007} here and rerun our model using the
H-chondrite value Hf/W$=0.7$. The results are presented in Figure
\ref{f:Hf-W_simple_lowHfW}. The pebble accretion time-scale of 5 million years
is now consistent with equilibration degrees as low as $f_{\rm eq}=0.5$ and
$f_{\rm eq}=0.25$. The isotopic composition of Earth and Mars in their major
lithophile elements can be matched with a mixture of inner Solar System and
outer Solar System material
\citep{Trinquier+etal2009,Warren2011,Schiller+etal2018}. The starting value of
Hf/W for Earth and Mars was therefore likely between the H-chondritic value and
the carbonaceous chondrite value. Figures \ref{f:Hf-W_simple} and
\ref{f:Hf-W_simple_lowHfW} thus represent endmember solutions for pure inner or
outer Solar System compositions. We conclude overall that the metal-silicate
equilibration degree in the moon-forming giant impact should not be much below
the range of 25\%-50\%, in order for a 5-million-years accretion time-scale to
be in agreement with the Hf-W system.

\subsection{Mixing degree in the giant impact}

The very similar ratio of $^{182}$W/$^{184}$W measured on Earth and the Moon
speak for efficient mixing of impactor and target after the moon-forming impact
\citep{Touboul+etal2007}. Any material from the mantle of proto-Earth that
experienced equilibration with the impactor must have been efficiently mixed
with the material that formed the Moon, in order to explain the similar W
isotopic abundances. Alternatively, the value of $^{182}$W/$^{184}$W could have
been similar on Theia and proto-Earth.  However, such a correspondence would be
very unlikely in the traditional giant impact scenario for planetary
differentiation, since the two bodies would have experienced stochastic impact
histories. Mars has an elevated value of $\epsilon_{\rm W}$ relative to Earth,
which indicates very rapid core formation within a few Myr after the formation
of the Solar System \citep{DauphasPourmand2011}. Such an impactor would require
extensive equilibration and mixing in the impact to drive $\epsilon_{\rm W}$
towards zero for both the Earth and the Moon.

Alternatively, Earth and Theia could have both formed their cores after the
extinction of $^{182}$Hf, but then the Earth-Moon system should have remained
chondritic $\epsilon_{\rm W}$ ($=-1.9$) and the measured elevation is
unexplained. The slight offset of the $^{182}$W/$^{184}$W of the Moon compared
to Earth (at the 20 ppm level, or 0.2 $\epsilon$ units) could be a signature
of a later veneer of chondritic material that preferentially enriched the more
massive Earth compared to the Moon \citep{Touboul+etal2015,KruijerKleine2017}.

\cite{DahlStevenson2010} used fluid dynamical models of planetesimals impacting
a larger planet to show that the emulsification of the iron core of the impactor
becomes increasingly inefficient as the impactor mass increases. However, a
collision between two nearly equal-sized planets, particularly if not directly
head on, would lead to a large-scale mixing of the two bodies \citep{Canup2012}.
The core of the impactor is sheared into multiple smaller components during the
first stage of the collision and these smaller metal blobs subsequently undergo
efficient emulsification while sedimenting to the core of the combined planet
\citep{Deguen+etal2014,Genda+etal2017}. Laboratory experiments show that taking
into account the inertia of the impactor dramatically increases the degree of
mixing \citep{Landeau+etal2021}. \cite{Maas+etal2021} utilized convective magma
ocean models and concluded that the convection and rotation of the planet leads
to much larger fractions of equilibrated volume than previously estimated and
that the equilibrated volume fraction increases with increasing impactor mass.
Hence a large degree of equilibration between a planetary-mass impactor and
proto-Earth following a planetary instability seems supported by hydrodynamical
simulations of impactor core emulsification.

\section{Discussion and implications}

The outgassing of volatiles to form a thick atmosphere during the planetary
accretion phase is key to achieving interior differentiation of terrestrial
planets. Pebbles deposit their accretion energy at the planetary surface and
this energy radiates away through the hydrogen envelope attracted from the
protoplanetary disc relatively easily. However, the outgassed atmosphere of
mainly H$_2$O and CO$_2$ provides a strong thermal blanketing that traps the
accretion energy and heats the surface to well above the liquidus of the
silicates. Magma oceans thus form as a natural evolution step in the planet
formation process once the protoplanet reaches a mass above approximately 2\% of
an Earth mass. We nevertheless demonstrate that the differentiation threshold
mass has a weak dependency on the assumed opacity of the atmospheric
constituents. Future models using frequency-dependent radiative transfer with
realistic opacity tables will therefore be useful to pin down more precisely the
value of the threshold mass for differentiation.

The trapped accretion energy would only penetrate slowly downwards through the
magma ocean. However, a more important source of mantle heating comes from the
continuous separation of metal melt from silicate melt and sedimentation of the
metal melt to the bottom of the magma ocean. This releases enough gravitational
potential energy to rapidly expand the magma ocean all the way down to the metal
core.  Subsequent accreted pebble material experiences rapid separation of metal
from silicates at the top of the magma ocean and fast sedimentation to join the
core.

The consistency of Earth's Hf-W system with the rapid accretion time-scale
of ${<}10$ Myr followed by an additional major mass contribution from the
moon-forming impactor was highlighted by \cite{YuJacobsen2011}. We have
demonstrated here a match between the Hf-W system and a pebble accretion model
with accretion time-scale as short as 5 Myr, followed by a collision with a
planetary body of mass in the range between $0.3\,M_{\rm E}$ and $0.5\,M_{\rm
E}$, in agreement with high-mass and high-angular momentum models for the
moon-forming giant impact as well as the necessity for extensive mixing of
impactor and target \citep{Canup2012}. In this view, the moon-forming impact is
not the last of a series of giant impacts that formed Earth, but rather the
result of an instability in a primordial terrestrial planet system that
contained an additional planet between Earth and Mars.

We have nevertheless found that impactor masses consistent with the Hf-W system
depend strongly on the assumed initial ratio of Hf/W as well as on the degree of
equilibration between the core of the impactor and the mantle of proto-Earth. A
low value of Hf/W -- consistent with measurements of ordinary chondrites
\citep{NimmoKleine2007} -- allow impactor masses as low as $0.2\,M_{\rm E}$,
more akin to the canonical impact model \citep{CanupAsphaug2001}, or low an
equilibrium fraction as low as $f_{\rm eq}=0.25$.

The preservation of anomalous $^{182}$W isotope signals in Earth's deep mantle,
expressed by both excesses and deficits relative to the upper mantle composition
\citep[i.e.][]{Mundl+etal2017,Willbold+etal2011} has been suggested to imply
inefficient mixing between Earth and the moon-forming impactor and, hence,
speak against a planetary-mass impactor. However, the small $^{182}$W deficits
identified in modern deep-seated mantle plumes are thought to reflect
interaction with Earth's core whereas the $^{182}$W excesses measured in Archean
rocks \citep{Mundl-Petermeier+etal2020,Rizo+etal2019} are interpreted to reflect
the composition of Earth’s mantle prior to the addition of the late veneer
\citep{Willbold+etal2011,Tusch+etal2021}. As such, the preservation of $^{182}$W
heterogeneities in Earth's deep mantle does not provide evidence for limited
mixing during the moon-forming impact.

\begin{acknowledgements}

We thank an anonymous referee for carefully reading the three papers in this
series and for giving us many comments and questions that helped improve the
original manuscripts. We also thank the second referee of this paper for
constructive comments. A.J.\ acknowledges funding from the European Research
Foundation (ERC Consolidator Grant 724687-PLANETESYS), the Knut and Alice
Wallenberg Foundation (Wallenberg Scholar Grant 2019.0442), the Swedish Research
Council (Project Grant 2018-04867), the Danish National Research Foundation
(DNRF Chair Grant DNRF159) and the G\"oran Gustafsson Foundation.  M.B.\
acknowledges funding from the Carlsberg Foundation (CF18\_1105) and the European
Research Council (ERC Advanced Grant 833275-DEEPTIME). M.S.\ acknowledges
funding from Villum Fonden (grant number \#00025333) and the Carlsberg
Foundation (grant number CF20-0209). The computations were enabled by resources
provided by the Swedish National Infrastructure for Computing (SNIC), partially
funded by the Swedish Research Council through grant agreement no.\ 2020/5-387.

\end{acknowledgements}

\appendix

\section{Numerical scheme for atmospheric luminosity}

The planet possesses both an outgassed atmosphere (of mass $M_{\rm atm}$) and an
envelope attracted from the protoplanetary disc (of mass $M_{\rm env}$). The
goal is to find the luminosity of the planet as a function of the known surface
temperature and the disc pressure and temperature boundary conditions (for
embedded planets) or the envelope mass (for isolated planets). We assume that
the opacity at the disc temperature is $\kappa_{\rm d}$ and that it increases with
temperature as $T^\beta$ with $\beta=2$. This is only approximately true for the
opacity at temperatures below the sublimation temperature of dust, but the
radiative-convective boundary typically occurs when the temperature is well
below the sublimation temperature.

\subsection{Embedded planet}

We consider first a planet still embedded within the protoplanetary disc. Here
the heat is removed from the planet--atmosphere--envelope system by the
temperature boundary condition at the Hill radius. We therefore follow
\cite{PisoYoudin2014} and integrate inwards from the disc boundary condition at
the Hill radius. We define the quantities
\begin{eqnarray}
  \nabla_{\rm ad} &=& \frac{\gamma-1}{\gamma} \, , \\
  \nabla_\infty   &=& \frac{1}{4-\beta} \, , \\
  \nabla_{\rm d}  &=& \frac{3 \kappa_{\rm d} P_{\rm d}}{64 \pi G M \sigma_{\rm
  SB} T_{\rm d}^4} L \, .
\end{eqnarray}
Here $\gamma$ is the adiabatic index of the envelope gas, assumed here to be
$\gamma=1.4$ valid for molecular hydrogen. The luminosity of the planet is $L$
and $\kappa_{\rm d}$, $P_{\rm d}$ and $T_{\rm d}$ are the opacity, pressure and
temperature at the outer boundary. The temperature profile changes from
isothermal to adiabatic at the radiative-convective boundary $R_{\rm rcb}$.
The pressure at the radiative-convective radius is
\begin{equation}
  P_{\rm rcb} = P_{\rm d} \frac{\nabla_{\rm ad}/\nabla_{\rm d}-\nabla_{\rm
  ad}/\nabla_\infty}{1-\nabla_{\rm ad}/\nabla_\infty} \, .
\end{equation}
The temperature at the radiative-convective boundary follows from integration
from the Hill radius through the radiative envelope solution and becomes 
\begin{equation}
  T_{\rm rcb} = T_{\rm d} \left[ \frac{\nabla_{\rm d}}{\nabla_\infty} \left(
  \frac{P_{\rm rcb}}{P_{\rm d}}-1 \right) + 1 \right]^{1/(4-\beta)} \, .
\end{equation}
\cite{PisoYoudin2014} showed that the location of the radiative-convective
boundary is close to the Bondi radius $R_{\rm B} = G M / c_{\rm s}^2$ with
$c_{\rm s}$ denoting the sound speed at the temperature of the outer boundary,
\begin{equation}
  R_{\rm rcb} = \frac{R_{\rm B}}{\ln(P_{\rm rcb}/P_{\rm d})} \, .
\end{equation}
The temperature at the bottom of the envelope follows integration through
the adiabatic envelope,
\begin{equation}
  T_{\rm env} = T_{\rm rcb} + \left( \frac{G M}{R} - \frac{G M}{R_{\rm rcb}}
  \right) \frac{1}{c_{\rm p}} \, .
\end{equation}
From $T_{\rm rcb}$, $P_{\rm rcb}$ and $T_{\rm env}$ follows $P_{\rm env}$ by
the adiabatic connection $P \propto T^{\gamma/(\gamma-1)}$. All these quantities
are therefore given uniquely by the set $(P_{\rm d}, T_{\rm d}, L)$.
\begin{figure}
  \includegraphics[width=\linewidth]{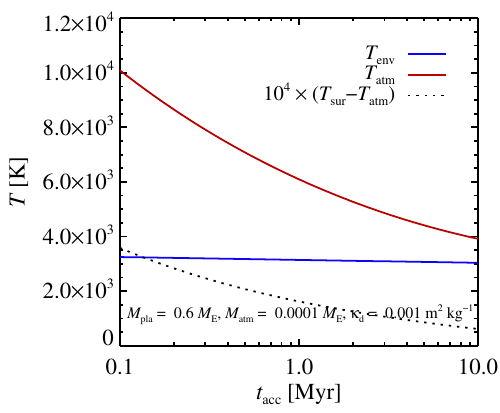}
  \caption{The envelope temperature, atmospheric bottom temperature
  and surface temperature difference as a function of the accretion time-scale
  (a measure of the luminosity through $L = G M \dot{M} /R$). The envelope
  temperature is close to 3,500 K for all values of the accretion time-scale.
  Matching this temperature at the top of the atmosphere leads to atmospheric
  temperatures above 10,000 K for short accretion times (high luminosities) and
  4,000 K for long accretion times (low luminosities).}
  \label{f:Tsur_Latmenv}
\end{figure}

This temperature and pressure must then match the temperature and pressure at
the top of the outgassed atmosphere. We know $P_{\rm atm}$ from the envelope
pressure and the atmospheric mass,
\begin{equation}
  P_{\rm atm} = \frac{g M_{\rm atm}}{4 \pi R^2} + P_{\rm env} \, ,
\end{equation}
under the assumption that the thickness of the atmosphere is much lower than the
planetary radius so that the gravitational acceleration $g$ can be assumed
constant. We can then test different values of $T_{\rm atm}$ (the temperature at
the bottom of the atmosphere) until the conditions at the top of the atmosphere,
where the pressure has fallen to $P = P_{\rm env}$, match the temperature and
pressure conditions at the bottom of the envelope. The luminosity is directly
related to the surface temperature $T_{\rm sur}$ and the temperature at the
bottom of the atmosphere $T_{\rm atm}$ through
\begin{equation}
  L_{\rm atm} = 4 \pi R_{\rm pla}^2 \sigma_{\rm SB} (T_{\rm sur}^4-T_{\rm
  atm}^4) \, .
  \label{eq:Latm}
\end{equation}
The atmosphere-envelope match is thus obtained at a unique luminosity $L_{\rm
atm}$ for a known surface temperature $T_{\rm sur}$.

The outgassed atmosphere consists of multiple chemical species, including water
with a tabulated equation of state. The atmospheric integration therefore has no
analytical solution. We can nevertheless test our algorithm considering an
outgassed atmosphere consisting entirely of CO$_2$ with an ideal-gas equation of
state and an adiabatic index of $\gamma=1.3$. The envelope temperature,
atmospheric temperature and surface temperature are shown in Figure
\ref{f:Tsur_Latmenv} for a planetary mass of $M = 0.6\,M_{\rm E}$, an atmosphere
mass of $M_{\rm atm} = 10^{-4} M_{\rm E}$ and a protoplanetary disc opacity of
$\kappa_{\rm d} = 0.001\,{\rm m^2\,kg^{-1}}$. The envelope temperature lies around $T
\approx 3,500\,{\rm K}$ for the considered range of accretion time-scales (which
define the luminosity). The surface temperature and atmospheric temperature are
indistinguishable, reflecting that the effective cooling temperature is much
lower than the surface temperature.

\subsection{Isolated planet}

After the dissipation of the protoplanetary disc, cooling is no longer taken
care of by the gas flow outside of the Hill radius. Instead, cooling is directly
achieved by loss of radiation from the photosphere. We assume in this case that
the hydrogen-helium envelope contracts to form a geometrically thin envelope. We
then calculate the photospheric temperature directly from the luminosity, $L = 4
\pi R^2 \sigma_{\rm SB} T_{\rm pho}^4$. The pressure at the photosphere follows
from integration of the hydrostatic equilibrium from $\tau=0$ to $\tau=2/3$,
which yields
\begin{equation}
  P_{\rm pho} = \frac{(2/3) g}{\kappa_{\rm d}} \, ,
\end{equation}
where $g$ is the gravitational acceleration at the surface and $\kappa_{\rm d}$
is the opacity of the envelope at the ambient temperature of the
now-dissipated protoplanetary disc (which we set to $\kappa_{\rm
d} = 0.1\,{\rm kg\,m^{-2}}$). The pressure at the bottom of the envelope follows
\begin{equation}
  P_{\rm env} = \frac{g M_{\rm env}}{4 \pi R^2} \, .
\end{equation}
The temperature at the bottom of the envelope is then calculated from the
adiabatic relation between the photosphere and the surface. We have compared
this approach to an algorithm that allows for the envelope to extend upwards to
regions of lower gravity and found no significant differences. The geometrically
thin assumption is therefore preferred, given its known analytical solution for
$P_{\rm env}$ as a function of $M_{\rm env}$. When the envelope has been removed
by the XUV irradiation from the star, the outgassed atmosphere is exposed
directly to vacuum cooling conditions.


\begin{thebibliography}{}
\bibitem[Armstrong et al.(2019)]{Armstrong+etal2019}
  Armstrong, K., Frost, D.~J., McCammon, C.~A., et al.\ 2019, Science, 365, 903
\bibitem[Badescu(2010)]{Badescu2010}
  Badescu, V.\ 2010, Central European Journal of Physics, 8, 463
\bibitem[Badro et al.(2015)]{Badro+etal2015}
  Badro, J., Brodholt, J.~P., Piet, H., et al.\ 2015, Proceedings of the
  National Academy of Science, 112, 12310
\bibitem[Bai \& Stone(2010)]{BaiStone2010}
  Bai, X.-N. \& Stone, J.~M.\ 2010, \apj, 722, 1437
\bibitem[Bitsch et al.(2015)]{Bitsch+etal2015}
  Bitsch, B., Johansen, A., Lambrechts, M., \& Morbidelli, A.\ 2015, \aap, 575,
  A28
\bibitem[Bouvier et al.(2018)]{Bouvier+etal2018}
  Bouvier, L.~C., Costa, M.~M., Connelly, J.~N., et al.\ 2018, \nat, 558, 568
\bibitem[Bower et al.(2022)]{Bower+etal2022}
  Bower, D.~J., Hakim, K., Sossi, P.~A., et al.\ 2022, The Planetary Science
  Journal, 3, 93
\bibitem[Brouwers et al.(2018)]{Brouwers+etal2018}
  Brouwers, M.~G., Vazan, A., \& Ormel, C.~W.\ 2018, \aap, 611, A65
\bibitem[Canup \& Asphaug(2001)]{CanupAsphaug2001}
  Canup, R.~M. \& Asphaug, E.\ 2001, \nat, 412, 708
\bibitem[Canup(2012)]{Canup2012}
  Canup, R.~M.\ 2012, Science, 338, 1052
\bibitem[Carlson et al.(2014)]{Carlson+etal2014}
  Carlson, R.~W., Garnero, E., Harrison, T.~M., et al.\ 2014, Annual Review of Earth and Planetary Sciences, 42, 151
\bibitem[Carrasco-Gonz{\'a}lez et al.(2019)]{Carrasco-Gonzalez+etal2019}
  Carrasco-Gonz{\'a}lez, C., Sierra, A., Flock, M., et al.\ 2019, \apj, 883, 71
\bibitem[Connelly \& Bizzarro(2016)]{ConnellyBizzarro2016}
  Connelly, J.~N. \& Bizzarro, M.\ 2016, Earth and Planetary Science Letters,
  452, 36
\bibitem[Dahl \& Stevenson(2010)]{DahlStevenson2010}
  Dahl, T.~W. \& Stevenson, D.~J.\ 2010, Earth and Planetary Science Letters, 295, 177
\bibitem[Dauphas \& Pourmand(2011)]{DauphasPourmand2011}
  Dauphas, N. \& Pourmand, A.\ 2011, \nat, 473, 489
\bibitem[Deguen et al.(2014)]{Deguen+etal2014}
  Deguen, R., Landeau, M., \& Olson, P.\ 2014, Earth and Planetary Science
  Letters, 391, 274
\bibitem[Deng et al.(2020)]{Deng+etal2020}
  Deng, J., Du, Z., Karki, B.~B., et al.\ 2020, Nature Communications, 11, 2007
\bibitem[Elkins-Tanton(2008)]{Elkins-Tanton2008}
  Elkins-Tanton, L.~T.\ 2008, Earth and Planetary Science Letters, 271, 181
\bibitem[Elkins-Tanton(2012)]{Elkins-Tanton2012}
  Elkins-Tanton, L.~T.\ 2012, Annual Review of Earth and Planetary Sciences, 40,
  113
\bibitem[Fischer et al.(2020)]{Fischer+etal2020}
  Fischer, R.~A., Cottrell, E., Hauri, E. Lee, K., \& Le Voyer, M.\ 2020,
  Proceedings of the National Academy of Sciences, 117, 8743
\bibitem[Gail \& Trieloff(2017)]{GailTrieloff2017}
  Gail, H.-P. \& Trieloff, M.\ 2017, \aap, 606, A16
\bibitem[Genda et al.(2017)]{Genda+etal2017}
  Genda, H., Brasser, R., \& Mojzsis, S.~J.\ 2017, Earth and Planetary Science
  Letters, 480, 25
\bibitem[Goldblatt et al.(2013)]{Goldblatt+etal2013}
  Goldblatt, C., Robinson, T.~D., Zahnle, K.~J., et al.\ 2013, Nature
  Geoscience, 6, 661
\bibitem[Grewal et al.(2019)]{Grewal+etal2019}
  Grewal, D.~S., Dasgupta, R., Holmes, A.~K., et al.\ 2019, \gca, 251, 87
\bibitem[Hamano et al.(2013)]{Hamano+etal2013}
  Hamano, K., Abe, Y., \& Genda, H.\ 2013, \nat, 497, 607
\bibitem[Hirschmann(2012)]{Hirschmann2012}
  Hirschmann, M.~M.\ 2012, Earth and Planetary Science Letters, 341, 48
\bibitem[Hirschmann(2021)]{Hirschmann2021}
  Hirschmann, M.~M.\ 2021, \gca, 313, 74
\bibitem[Hirschmann(2022)]{Hirschmann2022}
  Hirschmann, M.~M.\ 2022, \gca, 328, 221
\bibitem[Iacono-Marziano et al.(2012)]{Iacono-Marziano+etal2012}
  Iacono-Marziano, G., Morizet, Y., Le Trong, E., et al.\ 2012, \gca, 97, 1
\bibitem[Ingersoll(1969)]{Ingersoll1969}
  Ingersoll, A.~P.\ 1969, Journal of Atmospheric Sciences, 26, 1191
\bibitem[Jacobsen(2005)]{Jacobsen2005}
  Jacobsen, S.~B.\ 2005, Annual Review of Earth and Planetary Sciences, 33, 531
\bibitem[Jennings et al.(2021)]{Jennings+etal2021}
  Jennings, E.~S., Jacobson, S.~A., Rubie, D.~C., et al.\ 2021, \gca, 293, 40
\bibitem[Johansen et al.(2007)]{Johansen+etal2007}
  Johansen, A., Oishi, J.~S., Mac Low, M.-M., et al.\ 2007, \nat, 448, 1022
\bibitem[Johansen \& Lacerda(2010)]{JohansenLacerda2010}
  Johansen, A. \& Lacerda, P.\ 2010, \mnras, 404, 475
\bibitem[Johansen et al.(2015)]{Johansen+etal2015}
  Johansen, A., Mac Low, M.-M., Lacerda, P., et al.\ 2015, Science Advances, 1,
  1500109
\bibitem[Johansen \& Bitsch(2019)]{JohansenBitsch2019}
  Johansen, A. \& Bitsch, B.\ 2019, \aap, 631, A70
\bibitem[Johansen et al.(2021)]{Johansen+etal2021}
  Johansen, A., Ronnet, T., Bizzarro, M., Schiller, M., Lambrechts, M.,
  Nordlund, \AA, \& Lammer, H..\ 2021, Science Advancs, in press
\bibitem[Karman et al.(2019)]{Karman+etal2019}
  Karman, T., Gordon, I.~E., van der Avoird, A., et al.\ 2019, \icarus,
  328, 160. doi:10.1016/j.icarus.2019.02.034
\bibitem[Kleine \& Walker(2017)]{KleineWalker2017}
  Kleine, T. \& Walker, R.~J.\ 2017, Annual Review of Earth and Planetary
  Sciences, 45, 389
\bibitem[Kleine et al.(2009)]{Kleine+etal2009}
  Kleine, T., Touboul, M., Bourdon, B., et al.\ 2009, \gca, 73, 5150
\bibitem[Kokubo \& Ida(1998)]{KokuboIda1998}
  Kokubo, E. \& Ida, S.\ 1998, \icarus, 131, 171
\bibitem[Koll \& Cronin(2019)]{KollCronin2019}
  Koll, D.~D.~B. \& Cronin, T.~W.\ 2019, \apj, 881, 120
\bibitem[Kopparapu et al.(2013)]{Kopparapu+etal2013}
  Kopparapu, R.~K., Ramirez, R., Kasting, J.~F., et al.\ 2013, \apj, 765, 131
\bibitem[Kruijer \& Kleine(2017)]{KruijerKleine2017}
  Kruijer, T.~S. \& Kleine, T.\ 2017, Earth and Planetary Science Letters, 475,
  15
\bibitem[Kuramoto \& Matsui(1996)]{KuramotoMatsui1996}
  Kuramoto, K. \& Matsui, T.\ 1996, \jgr, 101, 14909
\bibitem[Lambrechts \& Johansen(2012)]{LambrechtsJohansen2012}
  Lambrechts, M., \& Johansen, A.\ 2012, \aap, 544, A32
\bibitem[Lambrechts et al.(2019)]{Lambrechts+etal2019}
  Lambrechts, M., Morbidelli, A., Jacobson, S.~A., et al.\ 2019, \aap, 627, A83
\bibitem[Landeau et al.(2021)]{Landeau+etal2021}
  Landeau, M., Deguen, R., Phillips, D., et al.\ 2021, Earth and Planetary
  Science Letters, 564, 116888
\bibitem[Leconte et al.(2013)]{Leconte+etal2013}
  Leconte, J., Forget, F., Charnay, B., et al.\ 2013, \nat, 504, 268
\bibitem[Li et al.(2020)]{Li+etal2020}
  Li, Y., Vo{\v{c}}adlo, L., Sun, T., et al.\ 2020, Nature Geoscience, 13, 453
\bibitem[Libourel et al.(2003)]{Libourel+etal2003}
  Libourel, G., Marty, B., \& Humbert, F.\ 2003, \gca, 67, 4123
\bibitem[Lichtenberg et al.(2021)]{Lichtenberg+etal2021}
  Lichtenberg, T., Bower, D.~J., Hammond, M., et al.\ 2021, Journal of
  Geophysical Research (Planets), 126, e06711
\bibitem[Liu(2019)]{Liu2019}
  Liu, H.~B.\ 2019, \apjl, 877, L22
\bibitem[Maas et al.(2021)]{Maas+etal2021}
  Maas, C., Manske, L., W{\"u}nnemann, K., et al.\ 2021, Earth and Planetary
  Science Letters, 554, 116680
\bibitem[Manger \& Klahr(2018)]{MangerKlahr2018}
  Manger, N. \& Klahr, H.\ 2018, \mnras, 480, 2125
\bibitem[Mann et al.(2009)]{Mann+etal2009}
  Mann, U., Frost, D.~J., \& Rubie, D.~C.\ 2009, \gca, 73, 7360
\bibitem[Matsui \& Abe(1986a)]{MatsuiAbe1986a}
  Matsui, T. \& Abe, Y.\ 1986, \nat, 319, 303
\bibitem[Matsui \& Abe(1986b)]{MatsuiAbe1986b}
  Matsui, T. \& Abe, Y.\ 1986, \nat, 322, 526
\bibitem[Misener \& Schlichting(2022)]{MisenerSchlichting2022}
  Misener, W. \& Schlichting, H.~E.\ 2022, \mnras, 514, 6025
\bibitem[Monteux et al.(2016)]{Monteux+etal2016}
  Monteux, J., Andrault, D., \& Samuel, H.\ 2016, Earth and Planetary Science
  Letters, 448, 140
\bibitem[Mundl et al.(2017)]{Mundl+etal2017}
  Mundl, A., Touboul, M., Jackson, M.~G., et al.\ 2017, Science, 356, 66
\bibitem[Mundl-Petermeier et al.(2020)]{Mundl-Petermeier+etal2020}
  Mundl-Petermeier, A., Walker, R.~J., Fischer, R.~A., et al.\ 2020, \gca, 271,
  194
\bibitem[Nakajima et al.(1992)]{Nakajima+etal1992}
  Nakajima, S., Hayashi, Y.-Y., \& Abe, Y.\ 1992, Journal of Atmospheric
  Sciences, 49, 2256
\bibitem[Nakano et al.(2003)]{Nakano+etal2003}
  Nakano, H., Kouchi, A., Tachibana, S., et al.\ 2003, \apj, 592, 1252
\bibitem[Nesvorn{\'y} et al.(2019)]{Nesvorny+etal2019}
  Nesvorn{\'y}, D., Li, R., Youdin, A.~N., et al.\ 2019, Nature Astronomy, 3,
  808
\bibitem[Nimmo \& Kleine(2007)]{NimmoKleine2007}
  Nimmo, F. \& Kleine, T.\ 2007, \icarus, 191, 497
\bibitem[Ormel \& Klahr(2010)]{OrmelKlahr2010}
  Ormel, C.~W., \& Klahr, H.~H.\ 2010, \aap, 520, A43
\bibitem[Ortenzi et al.(2020)]{Ortenzi+etal2020}
  Ortenzi, G., Noack, L., Sohl, F., et al.\ 2020, Scientific Reports, 10, 10907
\bibitem[Pack et al.(2011)]{Pack+etal2011}
  Pack, A., Vogel, I., Rollion-Bard, C., et al.\ 2011, Meteoritics \& Planetary
  Science, 46, 1470
\bibitem[Papale(1997)]{Papale1997}
  Papale, P.\ 1997, Contributions to Mineralogy and Petrology, 126, 237
\bibitem[Piso \& Youdin(2014)]{PisoYoudin2014}
  Piso, A.-M.~A. \& Youdin, A.~N.\ 2014, \apj, 786, 21
\bibitem[Olson et al.(2022)]{Olson+etal2022}
  Olson, P., Sharp, Z., \& Garai, S.\ 2022, Earth and Planetary Science Letters,
  587, 117537
\bibitem[Quarles \& Lissauer(2015)]{QuarlesLissauer2015}
  Quarles, B.~L. \& Lissauer, J.~J.\ 2015, \icarus, 248, 318
\bibitem[Rai \& van Westrenen(2013)]{RaivanWestrenen2013}
  Rai, N. \& Westrenen, W.\ 2013, Journal of Geophysical Research (Planets),
  118, 1195
\bibitem[Raymond et al.(2004)]{Raymond+etal2004}
  Raymond, S.~N., Quinn, T., \& Lunine, J.~I.\ 2004, \icarus, 168, 1
\bibitem[Righter \& Drake(1999)]{RighterDrake1999}
  Righter, K. \& Drake, M.~J.\ 1999, Earth and Planetary Science Letters, 171,
  383
\bibitem[Righter \& Chabot(2011)]{RighterChabot2011}
  Righter, K. \& Chabot, N.~L.\ 2011, Meteoritics and Planetary Science, 46,
  157
\bibitem[Righter \& O'Brien(2011)]{RighterO'Brien2011}
  Righter, K. \& O'Brien, D.~P.\ 2011, Proceedings of the National Academy of
  Science, 108, 19165
\bibitem[Rizo et al.(2019)]{Rizo+etal2019}
  Rizo, H., Andrault, D., Bennett, N.~R., et al.\ 2019, Geochem.\ Perspect.\
  Letter.\, 11, 6
\bibitem[Robinson \& Taylor(2001)]{RobinsonTaylor2001}
  Robinson, M.~S. \& Taylor, G.~J.\ 2001, Meteoritics and Planetary Science, 36,
  841
\bibitem[Rubie et al.(2003)]{Rubie+etal2003}
  Rubie, D.~C., Melosh, H.~J., Reid, J.~E., et al.\ 2003, Earth and Planetary
  Science Letters, 205, 239
\bibitem[Rubie et al.(2015a)]{Rubie+etal2015a}
  Rubie, D.~C., Jacobson, S.~A., Morbidelli, A., et al.\ 2015a, \icarus, 248, 89
\bibitem[Rubie et al.(2015b)]{Rubie+etal2015b}
  Rubie, D.~C., Nimmo, F., Melosh, H.~J., et al.\ 2015b, Treatise on Geophysics,
  Second Edition
\bibitem[Rudge et al.(2010)]{Rudge+etal2010}
  Rudge, J.~F., Kleine, T., \& Bourdon, B.\ 2010, Nature Geoscience, 3, 439
\bibitem[Salvador et al.(2017)]{Salvador+etal2017}
  Salvador, A., Massol, H., Davaille, A., et al.\ 2017, Journal of Geophysical
  Research (Planets), 122, 1458
\bibitem[Schaefer \& Fegley(2004)]{SchaeferFegley2004}
  Schaefer, L. \& Fegley, B.\ 2004, \icarus, 169, 216
\bibitem[Schaefer \& Fegley(2010)]{SchaeferFegley2010}
  Schaefer, L. \& Fegley, B.\ 2010, \icarus, 208, 438
\bibitem[Schaefer \& Elkins-Tanton(2018)]{SchaeferElkins-Tanton2018}
  Schaefer, L. \& Elkins-Tanton, L.~T.\ 2018, Philosophical Transactions of the
  Royal Society of London Series A, 376, 20180109
\bibitem[Sch{\"a}fer et al.(2020)]{Schaefer+etal2020}
  Sch{\"a}fer, U., Johansen, A., \& Banerjee, R.\ 2020, \aap, 635, A190
\bibitem[Schiller et al.(2018)]{Schiller+etal2018}
  Schiller, M., Bizzarro, M., \& Fernandes, V.~A.\ 2018, \nat, 555, 507
\bibitem[Schiller et al.(2020)]{Schiller+etal2020}
  Schiller, M., Bizzarro, M., \& Siebert, J.\ 2020, Science Advances, 6,
  eaay7604
\bibitem[Simon et al.(2016)]{Simon+etal2016}
  Simon, J.~B., Armitage, P.~J., Li, R., et al.\ 2016, \apj, 822, 55
\bibitem[Sossi et al.(2020)]{Sossi+etal2020}
  Sossi, P.~A., Burnham, A.~D., Badro, J., et al.\ 2020, Science Advances, 6,
  eabd1387
\bibitem[St{\"o}kl et al.(2016)]{Stokl+etal2016}
  St{\"o}kl, A., Dorfi, E.~A., Johnstone, C.~P., et al.\ 2016, \apj, 825, 86
\bibitem[Stoll \& Kley(2016)]{StollKley2016}
  Stoll, M.~H.~R. \& Kley, W.\ 2016, \aap, 594, A57
\bibitem[Tanaka \& Tsukamoto(2019)]{TanakaTsukamoto2019}
  Tanaka, Y.~A., \& Tsukamoto, Y.\ 2019, \mnras, 484, 1574
\bibitem[Touboul et al.(2007)]{Touboul+etal2007}
  Touboul, M., Kleine, T., Bourdon, B., et al.\ 2007, \nat, 450, 1206
\bibitem[Touboul et al.(2015)]{Touboul+etal2015}
  Touboul, M., Puchtel, I.~S., \& Walker, R.~J.\ 2015, \nat, 520, 530
\bibitem[Trinquier et al.(2009)]{Trinquier+etal2009}
  Trinquier, A., Elliott, T., Ulfbeck, D., et al.\ 2009, Science, 324, 374
\bibitem[Tr{\o}nnes et al.(2019)]{Tronnes+etal2019}
  Tr{\o}nnes, R.~G., Baron, M.~A., Eigenmann, K.~R., et al.\ 2019,
  Tectonophysics, 760, 165
\bibitem[Tusch et al.(2021)]{Tusch+etal2021}
  Tusch, J., M{\"u}nker, C., Hasenstab, E., et al.\ 2021, Proceedings of the
  National Academy of Science, 118, 2012626118
\bibitem[Tychoniec et al.(2018)]{Tychoniec+etal2018}
  Tychoniec, {\L}., Tobin, J.~J., Karska, A., et al.\ 2018, \apjs, 238, 19
\bibitem[Wade \& Wood(2005)]{WadeWood2005}
  Wade, J. \& Wood, B.~J.\ 2005, Earth and Planetary Science Letters, 236, 78
\bibitem[Walter \& Cottrell(2013)]{WalterCottrell2013}
  Walter, M.~J. \& Cottrell, E.\ 2013, Earth and Planetary Science Letters, 365,
  165
\bibitem[Warren(2011)]{Warren2011}
  Warren, P.~H.\ 2011, Earth and Planetary Science Letters, 311, 93
\bibitem[Weiss \& Elkins-Tanton(2013)]{WeissElkins-Tanton2013}
  Weiss, B.~P. \& Elkins-Tanton, L.~T.\ 2013, Annual Review of Earth and
  Planetary Sciences, 41, 529
\bibitem[Willbold et al.(2011)]{Willbold+etal2011}
  Willbold, M., Elliott, T., \& Moorbath, S.\ 2011, \nat, 477, 195
\bibitem[Williams \& Mukhopadhyay(2019)]{WilliamsMukhopadhyay2019}
  Williams, C.~D. \& Mukhopadhyay, S.\ 2019, \nat, 565, 78
\bibitem[Wordsworth(2015)]{Wordsworth2015}
  Wordsworth, R.\ 2015, \apj, 806, 180
\bibitem[Wordsworth et al.(2021)]{Wordsworth+etal2021}
  Wordsworth, R., Knoll, A.~H., Hurowitz, J., et al.\ 2021, Nature Geoscience,
  14, 127
\bibitem[Wu et al.(2018)]{Wu+etal2018}
  Wu, J., Desch, S.~J., Schaefer, L., et al.\ 2018, Journal of Geophysical
  Research (Planets), 123, 2691
\bibitem[Xiao \& Stixrude(2018)]{XiaoStixrude2018}
  Xiao, B.-C. \& Stixrude, L.\ 2018, PNAS, 115, 5371
\bibitem[Yang et al.(2018)]{Yang+etal2018}
  Yang, C.-C., Mac Low, M.-M., \& Johansen, A.\ 2018, \apj, 868, 27
\bibitem[Yin et al.(2002)]{Yin+etal2002}
  Yin, Q., Jacobsen, S.~B., Yamashita, K., et al.\ 2002, \nat, 418, 949
\bibitem[Yu \& Jacobsen(2011)]{YuJacobsen2011}
  Yu, G. \& Jacobsen, S.~B.\ 2011, Proceedings of the National Academy of Science, 108, 17604
\bibitem[Zhu et al.(2019)]{Zhu+etal2019}
  Zhu, Z., Zhang, S., Jiang, Y.-F., et al.\ 2019, \apjl, 877, L18
\bibitem[Zhu et al.(2022)]{Zhu+etal2022}
  Zhu, K., Schiller, M., Pan, L., et al.\ 2022, Science Advances, 8, eabp8415

\end{thebibliography}
\end{document}